\documentclass[english,a4paper,11pt]{article}

\usepackage[english]{babel}
\usepackage{graphicx,epsfig,epic} 
\usepackage{pifont}      

\usepackage{pict2e}
\usepackage{amssymb}

\font\tenbb=msbm10 at 12pt

\def\rR{\hbox{\tenbb R}}

\def\esp{\vskip .6cm}
\def\pesp{\vskip .3cm}

\def\di{\displaystyle}

\newsavebox{\fmbox}
\newenvironment{fmpage}[1]
         {\begin{lrbox}{\fmbox}\begin{minipage}{#1}}
         {\end{minipage}\end{lrbox}\fbox{\usebox{\fmbox}}}

\begin{document}

\title {Paths of Least Time for Quantum Scale and a New Geometrical Interpretation of Light Diffraction}

\author{\small F. BEN ADDA{\footnote{New York Institute of Technology.
Email: f$\_$benaddafr@yahoo.fr}}}

\maketitle

\begin{abstract}
In this paper, a geometrical interpretation of light diffraction is given using an infinity of fluctuating geodesics for the small scale (the quantum scale) that represent paths of least time in an homogeneous space. Without using the wave theory, we provide a geometrical explanation of the deviation of light's overall direction from rectilinear when light encounters edges, apertures and screens.
\end{abstract}

{\footnotesize Pacs: 02.70.-c, 42.25.Hz}

\section{Introduction}
The quest for the comprehension of the real nature and properties of light has involved, for centuries, a big number of scientists and researchers such as Ibn Alhaitham, Descartes, Newton, Grimaldi, Young, Huygens, Fresnel, Maxwell, Hertz, R\"{o}ntgen, Von Laue, Planck, Einstein, Bohr, Compton, De Broglie, Davisson, Paget Thomson, Schr\"{o}dinger, Dirac, Tomonaga, Schwinger, Feynman, Dyson and  others (\cite{AA}, \cite{BN}, \cite{CA}, \cite{DC}, \cite{DB}, \cite{DR}, \cite{DI},  \cite{EA}, \cite{EG}, \cite{FA1}, \cite{FA2}, \cite{GO}, \cite{GR}, \cite{HH}, \cite{HD}, \cite{HC}, \cite{IH},  \cite{MC},  \cite{N1}, \cite{N2}, \cite{PM}, \cite{SA1}, \cite{SA2}, \cite{SC}, \cite{SS}, \cite{TG}, \cite{TO}, \cite{VL}, \cite{TY1}, \cite{TY2}). The progress of research in the $19^{th}$ century has shaped our way of thinking about the nature of light. The significant progress of the quantum theory based on experimental results and observations for the small scale world dictate our understanding of the nature of light and influence the way one can approach the photon's motion and nature. \pesp

The indeterminacy based on the Heisenberg's uncertainty principle imposes the non consideration of paths and trajectories for elementary particles and photons. This difficulty is optimized by the use of density of probabilities of particle position, where the state of an elementary particle is entirely given by the wave function (a unit vector in Hilbert space) that verifies Schr\"{o}dinger equation, meanwhile the wave associated to the photons verifies the electromagnetic wave equation.\pesp

 Providing a geometrical explanation of the light diffraction passes thought trajectories and the principle of least time. In order to be in conformity with the indeterminacy of the fundamental principle of quantum mechanics (Heisenberg's uncertainty), we will use in this work an infinity of paths of least time for the photon's motion that minimize the total time needed for the photon to travel between two locations without determination of a particular path. We will use an infinity of possible paths (that have the same length) for the photon to travel between the two distant locations.\pesp

 If a quantum physical system in a given space-time is in free motion from one location to another following the path that requires the shortest time, then the geometry of the space-time could dictate the behavior of the physical system in following the curve that minimizes the total time needed to travel between the two locations (this concept is used in cosmology with the so-called cosmological redshift: the light wavelength increases as it propagates through an expanding space by the same amount (\cite{HE}), which means that the expansion of the space affects the wavelength). If there exists some "interaction" between the expanding space and light that makes its wavelength increases by the same amount as it propagates through the expanding space, then the light motion must be shaped by the space characteristics (such as geodesics and paths of least time).\pesp

 Interesting geodesics are found in the simulation of an expanding space-time \cite{BP}, that expands via expansion of its basic elements, where the path of least time is found to be given by an infinity of fluctuating geodesics directed by a geodesics axis that indicates the overall displacement, and minimizes the total time needed to travel between two distant locations in an homogeneous space. The use of these geodesics has led to reproduce an interference pattern similar to the interference pattern observed in Young's double-slit experiment under the assumption that all possible geodesics of the physical system, that represent paths of least time and that pass through an opening, flare out (diffract) and cover an angle beyond the slit (see \cite{BF}). The deviation of the physical system's geodesics direction from rectilinear, when it passes through a narrow slit and creates the diffraction, is explained within this paper, which will complete the geometrical explanation of interference pattern of Young's double-slit experiment introduced in (\cite{BF}) without using the wave theory.

\section{Prototype of Space-Time Fluctuating Geodesics}
\subsection{Fluctuating Geodesics in the Plane}
Let us consider a physical system in a free motion between two distant locations A and B, in a given homogeneous space-time, following the path that requires the shortest time, and let us assume that the space-time geodesics are defined (see \cite{BP} for more details) by $\varphi$ given by
\begin{equation}\label{Geod}
\varphi(x)=\sum_{i=0}^{N-1}g_{i}(x), \qquad \hbox{for}\quad N\geq1
\end{equation}
where
\begin{equation}\label{SubGeod}
g_{i}(x)=\left\{
             \begin{array}{ll}
               \varphi_{i}(x) & \hbox{for}\quad x\in [2ir,2(i+1)r]\\
               0 & \hbox{for}\quad x\not\in [2ir,2(i+1)r]
             \end{array}
           \right.
\end{equation}
and
\begin{eqnarray}
  \nonumber \varphi_{i} :\ [2ir,2(i+1)r]&\longrightarrow& \rR \\
  x &\longmapsto & \varphi_{i}(x)\ =\ (-1)^i\ \sqrt{r^2 - \Big(x-(2i+1)r\Big)^2},
\end{eqnarray}
that verifies:\pesp

i) for $i=0,\dots, N-1$, the graph of $\varphi_{i}$ represents the geodesic between two antipodal points on the circle of center
$C_{i}=((2i+1)r, 0)$ and radius $r$;\pesp

ii) for $i=0,\dots, N-1$,\  $\varphi_{i}$ is continuous on the closed interval \quad $[2ir,2(i+1)r]$;\pesp

iii) for $i=0,\dots, N-1$,\  $\varphi_{i}$ is differentiable on the open interval \quad $]2ir,2(i+1)r[$;\pesp

iv) for $i=0,\dots, N-1$,\  $\varphi_{i}$ is not differentiable at the points\quad $x_{i}=2ir$\ and\ $x_{i+1}=2(i+1)r$.\pesp

The graph of these geodesics is illustrated for a given $r$ in Fig.1.

\vskip1cm
\setlength{\unitlength}{1cm}
\begin{center}
\begin{fmpage}{14.5cm}
\begin{picture}(5,5)
\put(1,2){\vector(1,0){13.5}}
\put(1,2){\vector(0,1){3}}
\put(0.7,1.5){$A$}
\put(14,1.5){$B$}
\put(1.5,2){\vector(1,1){.4}}
\put(0.7,2){$0$}
\put(1.3, 4.5){y}
\put(14,2.5){$x$}
\put(6.4,3.2){$4r$}
\put(5.5,3){\vector(1,0){2}}
\put(7.5,3){\vector(-1,0){2}}
\linethickness{.4mm}
\put(1.5,2){\oval[1](1,1)[t]}
\put(2.5,2){\oval[1](1,1)[b]}
\put(3.5,2){\oval[1](1,1)[t]}
\put(4.5,2){\oval[1](1,1)[b]}
\put(5.5,2){\oval[1](1,1)[t]}
\put(6.5,2){\oval[1](1,1)[b]}

\put(7.5,2){\oval[1](1,1)[t]}
\put(8.5,2){\oval[1](1,1)[b]}
\put(9.5,2){\oval[1](1,1)[t]}
\put(10.5,2){\oval[1](1,1)[b]}
\put(11.5,2){\oval[1](1,1)[t]}
\put(12.5,2){\oval[1](1,1)[b]}
\put(13.5,2){\oval[1](1,1)[t]}
\put(1.5,2){\circle*{.05}}
\put(1.4,2.2){$r$}
\linethickness{.02mm}
\put(1.5,2){\oval[1](1,1)[b]}
\put(2.5,2){\oval[1](1,1)[t]}
\put(3.5,2){\oval[1](1,1)[b]}
\put(4.5,2){\oval[1](1,1)[t]}
\put(5.5,2){\oval[1](1,1)[b]}
\put(6.5,2){\oval[1](1,1)[t]}

\put(7.5,2){\oval[1](1,1)[b]}
\put(8.5,2){\oval[1](1,1)[t]}
\put(9.5,2){\oval[1](1,1)[b]}
\put(10.5,2){\oval[1](1,1)[t]}
\put(11.5,2){\oval[1](1,1)[b]}
\put(12.5,2){\oval[1](1,1)[t]}
\put(13.5,2){\oval[1](1,1)[b]}
\put(0,-1){\shortstack[l]{\small{Figure 1:} \footnotesize Graph of the geodesics $\varphi(x)$ and $-\varphi(x)$ in 2D for $N=13$ and radius $r=5\ mm$ between\\ \footnotesize two points A(0,0) and B(13,0). The graph of $\varphi(x)$ is black and the graph of $-\varphi(x)$ is grey.}}
\end{picture}
\end{fmpage}
\end{center}
\vskip1cm

Based on property i), if $\varphi_{i}(x)$ represents the path that requires the shortest time between two successive antipodal points, then $-\varphi_{i}(x)$ is also a path of least time between the same antipodal points, which means that if a physical system is moving between two distant locations $A=(0,0)$ and $B=(x,\varphi_{N-1}(x))$ for $x\in]2(N-1)r,2Nr]$, the path of least time can be the graph of $\varphi(x)$, or in general the graph of the geodesics
\begin{equation}\label{GenGeod}
\Psi(x)=\left\{
  \begin{array}{ll}
\di\sum_{i=0}^{N-1}\pm g_{i}(x),  & \hbox{if} \quad x=2(i+1)r; \\
\di\sum_{i=0}^{N-2}\pm g_{i}(x)+ g_{N-1}(x), & \hbox{if} \quad x\in]2ir,2(i+1)r[.
\end{array}
\right.
\end{equation}
where $g_i$ is defined by (\ref{SubGeod}). Indeed, there exist two opposite geodesics between two antipodal points on the circle of center
$C_{i}=((2i+1)r, 0)$ and radius $r$ for all $i=0,...,N-1$, which means that between the two distant locations $A=(0,0)$ and $B=(x,\varphi_{N-1}(x))$ for $x\in]2(N-1)r,2Nr]$ there exist
\begin{equation}\label{least}
\left\{
  \begin{array}{ll}
    2^N\quad\hbox{geodesics}, & \hbox{if} \quad x=2(i+1)r; \\
    2^{N-1}\quad\hbox{geodesics}, & \hbox{if} \quad x\in]2ir,2(i+1)r[.
  \end{array}
\right.
\end{equation}
that represent paths of least time between the two locations $A$ and $B$. Any physical system that follows the path of least time given by (\ref{GenGeod}) in the plane will have $2^{N}$ or $2^{N-1}$ possibilities, then it is impossible to predict from which path the physical system will pass through.

\subsection{Fluctuating Geodesics in the Space}\label{Fluct}

To obtain all possible geodesics between two distant locations in three dimensions using the geodesic (\ref{Geod}), we use a rotation of the graph of the function (\ref{Geod}) about the $x$-axis (see Fig.2). Indeed, if
we denote the graph of $\varphi$ by
\begin{equation}
G_\varphi=\{(x,y,z)\in\rR^3\ /\ y=\varphi(x)\ , z=0\ \}
\end{equation}
then the graph of all geodesics in three dimensions is given by
\begin{equation}\label{Rot}
G_{R_\theta(\varphi)}=\Big\{(x',y',z')\in \rR^3\ /\ \left(
                                               \begin{array}{c}
                                                 x' \\
                                                 y' \\
                                                 z'\\
                                               \end{array}
                                             \right)=R_\theta \left(
                                                                \begin{array}{c}
                                                                  x \\
                                                                  y \\
                                                                  z \\
                                                                \end{array}
                                                              \right)
                                             ,\hbox{with}\ \left(
                                                                           \begin{array}{c}
                                                                             x \\
                                                                             y \\
                                                                             z \\
                                                                           \end{array}
                                                                         \right)\in G_\varphi
\ \Big\}
\end{equation}
where $R_\theta$ is the rotation about the $x$-axis of angle $\theta\in [0,2\pi]$ given by
\begin{equation}
R_\theta=\left(
  \begin{array}{ccc}
    1 & 0 & 0 \\
    0 & \cos\theta & -\sin\theta \\
    0 & \sin\theta & \cos\theta \\
  \end{array}
\right)
\end{equation}

\pesp
\setlength{\unitlength}{1cm}
\begin{center}
\begin{fmpage}{14.5cm}
\begin{picture}(5,5)
\put(1,2){\vector(1,0){13.5}}
\put(0.7,1.2){$A$}
\put(14,1.2){$B$}
\put(1,2){\vector(0,1){3}}
\put(1,2){\vector(-1,-1){1}}
\put(14,2.5){$x$}
\put(6.4,3.2){$4r$}
\put(5.5,3){\vector(1,0){2}}
\put(1.5,4.5){$y$}
\put(7.5,3){\vector(-1,0){2}}
\put(0,0.5){$z$}
\linethickness{.1mm}
\put(0,0){
\put(1.5,2){\oval[1](1,1)[t]}
\qbezier(1,2)(1,2.4)(1.5,2.4)
\qbezier(1.5,2.4)(2,2.4)(2,2)
\qbezier(1,2)(1,2.3)(1.5,2.3)
\qbezier(1.5,2.3)(2,2.3)(2,2)
\qbezier(1,2)(1,2.2)(1.5,2.2)
\qbezier(1.5,2.2)(2,2.2)(2,2)
\qbezier(1,2)(1,2.1)(1.5,2.1)
\qbezier(1.5,2.1)(2,2.1)(2,2)
\put(1.5,2){\oval[1](1,1)[b]}
\qbezier(1,2)(1,1.6)(1.5,1.6)
\qbezier(1.5,1.6)(2,1.6)(2,2)
\qbezier(1,2)(1,1.7)(1.5,1.7)
\qbezier(1.5,1.7)(2,1.7)(2,2)
\qbezier(1,2)(1,1.8)(1.5,1.8)
\qbezier(1.5,1.8)(2,1.8)(2,2)
\qbezier(1,2)(1,1.9)(1.5,1.9)
\qbezier(1.5,1.9)(2,1.9)(2,2)
}
\put(1,0){
\put(1.5,2){\oval[1](1,1)[t]}
\qbezier(1,2)(1,2.4)(1.5,2.4)
\qbezier(1.5,2.4)(2,2.4)(2,2)
\qbezier(1,2)(1,2.3)(1.5,2.3)
\qbezier(1.5,2.3)(2,2.3)(2,2)
\qbezier(1,2)(1,2.2)(1.5,2.2)
\qbezier(1.5,2.2)(2,2.2)(2,2)
\qbezier(1,2)(1,2.1)(1.5,2.1)
\qbezier(1.5,2.1)(2,2.1)(2,2)
\put(1.5,2){\oval[1](1,1)[b]}
\qbezier(1,2)(1,1.6)(1.5,1.6)
\qbezier(1.5,1.6)(2,1.6)(2,2)
\qbezier(1,2)(1,1.7)(1.5,1.7)
\qbezier(1.5,1.7)(2,1.7)(2,2)
\qbezier(1,2)(1,1.8)(1.5,1.8)
\qbezier(1.5,1.8)(2,1.8)(2,2)
\qbezier(1,2)(1,1.9)(1.5,1.9)
\qbezier(1.5,1.9)(2,1.9)(2,2)
}

\put(2,0){
\put(1.5,2){\oval[1](1,1)[t]}
\qbezier(1,2)(1,2.4)(1.5,2.4)
\qbezier(1.5,2.4)(2,2.4)(2,2)
\qbezier(1,2)(1,2.3)(1.5,2.3)
\qbezier(1.5,2.3)(2,2.3)(2,2)
\qbezier(1,2)(1,2.2)(1.5,2.2)
\qbezier(1.5,2.2)(2,2.2)(2,2)
\qbezier(1,2)(1,2.1)(1.5,2.1)
\qbezier(1.5,2.1)(2,2.1)(2,2)
\put(1.5,2){\oval[1](1,1)[b]}
\qbezier(1,2)(1,1.6)(1.5,1.6)
\qbezier(1.5,1.6)(2,1.6)(2,2)
\qbezier(1,2)(1,1.7)(1.5,1.7)
\qbezier(1.5,1.7)(2,1.7)(2,2)
\qbezier(1,2)(1,1.8)(1.5,1.8)
\qbezier(1.5,1.8)(2,1.8)(2,2)
\qbezier(1,2)(1,1.9)(1.5,1.9)
\qbezier(1.5,1.9)(2,1.9)(2,2)
}
\put(3,0){
\put(1.5,2){\oval[1](1,1)[t]}
\qbezier(1,2)(1,2.4)(1.5,2.4)
\qbezier(1.5,2.4)(2,2.4)(2,2)
\qbezier(1,2)(1,2.3)(1.5,2.3)
\qbezier(1.5,2.3)(2,2.3)(2,2)
\qbezier(1,2)(1,2.2)(1.5,2.2)
\qbezier(1.5,2.2)(2,2.2)(2,2)
\qbezier(1,2)(1,2.1)(1.5,2.1)
\qbezier(1.5,2.1)(2,2.1)(2,2)
\put(1.5,2){\oval[1](1,1)[b]}
\qbezier(1,2)(1,1.6)(1.5,1.6)
\qbezier(1.5,1.6)(2,1.6)(2,2)
\qbezier(1,2)(1,1.7)(1.5,1.7)
\qbezier(1.5,1.7)(2,1.7)(2,2)
\qbezier(1,2)(1,1.8)(1.5,1.8)
\qbezier(1.5,1.8)(2,1.8)(2,2)
\qbezier(1,2)(1,1.9)(1.5,1.9)
\qbezier(1.5,1.9)(2,1.9)(2,2)
}
\put(4,0){
\put(1.5,2){\oval[1](1,1)[t]}
\qbezier(1,2)(1,2.4)(1.5,2.4)
\qbezier(1.5,2.4)(2,2.4)(2,2)
\qbezier(1,2)(1,2.3)(1.5,2.3)
\qbezier(1.5,2.3)(2,2.3)(2,2)
\qbezier(1,2)(1,2.2)(1.5,2.2)
\qbezier(1.5,2.2)(2,2.2)(2,2)
\qbezier(1,2)(1,2.1)(1.5,2.1)
\qbezier(1.5,2.1)(2,2.1)(2,2)
\put(1.5,2){\oval[1](1,1)[b]}
\qbezier(1,2)(1,1.6)(1.5,1.6)
\qbezier(1.5,1.6)(2,1.6)(2,2)
\qbezier(1,2)(1,1.7)(1.5,1.7)
\qbezier(1.5,1.7)(2,1.7)(2,2)
\qbezier(1,2)(1,1.8)(1.5,1.8)
\qbezier(1.5,1.8)(2,1.8)(2,2)
\qbezier(1,2)(1,1.9)(1.5,1.9)
\qbezier(1.5,1.9)(2,1.9)(2,2)
}
\put(5,0){
\put(1.5,2){\oval[1](1,1)[t]}
\qbezier(1,2)(1,2.4)(1.5,2.4)
\qbezier(1.5,2.4)(2,2.4)(2,2)
\qbezier(1,2)(1,2.3)(1.5,2.3)
\qbezier(1.5,2.3)(2,2.3)(2,2)
\qbezier(1,2)(1,2.2)(1.5,2.2)
\qbezier(1.5,2.2)(2,2.2)(2,2)
\qbezier(1,2)(1,2.1)(1.5,2.1)
\qbezier(1.5,2.1)(2,2.1)(2,2)
\put(1.5,2){\oval[1](1,1)[b]}
\qbezier(1,2)(1,1.6)(1.5,1.6)
\qbezier(1.5,1.6)(2,1.6)(2,2)
\qbezier(1,2)(1,1.7)(1.5,1.7)
\qbezier(1.5,1.7)(2,1.7)(2,2)
\qbezier(1,2)(1,1.8)(1.5,1.8)
\qbezier(1.5,1.8)(2,1.8)(2,2)
\qbezier(1,2)(1,1.9)(1.5,1.9)
\qbezier(1.5,1.9)(2,1.9)(2,2)
}
\put(6,0){
\put(1.5,2){\oval[1](1,1)[t]}
\qbezier(1,2)(1,2.4)(1.5,2.4)
\qbezier(1.5,2.4)(2,2.4)(2,2)
\qbezier(1,2)(1,2.3)(1.5,2.3)
\qbezier(1.5,2.3)(2,2.3)(2,2)
\qbezier(1,2)(1,2.2)(1.5,2.2)
\qbezier(1.5,2.2)(2,2.2)(2,2)
\qbezier(1,2)(1,2.1)(1.5,2.1)
\qbezier(1.5,2.1)(2,2.1)(2,2)
\put(1.5,2){\oval[1](1,1)[b]}
\qbezier(1,2)(1,1.6)(1.5,1.6)
\qbezier(1.5,1.6)(2,1.6)(2,2)
\qbezier(1,2)(1,1.7)(1.5,1.7)
\qbezier(1.5,1.7)(2,1.7)(2,2)
\qbezier(1,2)(1,1.8)(1.5,1.8)
\qbezier(1.5,1.8)(2,1.8)(2,2)
\qbezier(1,2)(1,1.9)(1.5,1.9)
\qbezier(1.5,1.9)(2,1.9)(2,2)
}
\put(7,0){
\put(1.5,2){\oval[1](1,1)[t]}
\qbezier(1,2)(1,2.4)(1.5,2.4)
\qbezier(1.5,2.4)(2,2.4)(2,2)
\qbezier(1,2)(1,2.3)(1.5,2.3)
\qbezier(1.5,2.3)(2,2.3)(2,2)
\qbezier(1,2)(1,2.2)(1.5,2.2)
\qbezier(1.5,2.2)(2,2.2)(2,2)
\qbezier(1,2)(1,2.1)(1.5,2.1)
\qbezier(1.5,2.1)(2,2.1)(2,2)
\put(1.5,2){\oval[1](1,1)[b]}
\qbezier(1,2)(1,1.6)(1.5,1.6)
\qbezier(1.5,1.6)(2,1.6)(2,2)
\qbezier(1,2)(1,1.7)(1.5,1.7)
\qbezier(1.5,1.7)(2,1.7)(2,2)
\qbezier(1,2)(1,1.8)(1.5,1.8)
\qbezier(1.5,1.8)(2,1.8)(2,2)
\qbezier(1,2)(1,1.9)(1.5,1.9)
\qbezier(1.5,1.9)(2,1.9)(2,2)
}
\put(8,0){
\put(1.5,2){\oval[1](1,1)[t]}
\qbezier(1,2)(1,2.4)(1.5,2.4)
\qbezier(1.5,2.4)(2,2.4)(2,2)
\qbezier(1,2)(1,2.3)(1.5,2.3)
\qbezier(1.5,2.3)(2,2.3)(2,2)
\qbezier(1,2)(1,2.2)(1.5,2.2)
\qbezier(1.5,2.2)(2,2.2)(2,2)
\qbezier(1,2)(1,2.1)(1.5,2.1)
\qbezier(1.5,2.1)(2,2.1)(2,2)
\put(1.5,2){\oval[1](1,1)[b]}
\qbezier(1,2)(1,1.6)(1.5,1.6)
\qbezier(1.5,1.6)(2,1.6)(2,2)
\qbezier(1,2)(1,1.7)(1.5,1.7)
\qbezier(1.5,1.7)(2,1.7)(2,2)
\qbezier(1,2)(1,1.8)(1.5,1.8)
\qbezier(1.5,1.8)(2,1.8)(2,2)
\qbezier(1,2)(1,1.9)(1.5,1.9)
\qbezier(1.5,1.9)(2,1.9)(2,2)
}
\put(9,0){
\put(1.5,2){\oval[1](1,1)[t]}
\qbezier(1,2)(1,2.4)(1.5,2.4)
\qbezier(1.5,2.4)(2,2.4)(2,2)
\qbezier(1,2)(1,2.3)(1.5,2.3)
\qbezier(1.5,2.3)(2,2.3)(2,2)
\qbezier(1,2)(1,2.2)(1.5,2.2)
\qbezier(1.5,2.2)(2,2.2)(2,2)
\qbezier(1,2)(1,2.1)(1.5,2.1)
\qbezier(1.5,2.1)(2,2.1)(2,2)
\put(1.5,2){\oval[1](1,1)[b]}
\qbezier(1,2)(1,1.6)(1.5,1.6)
\qbezier(1.5,1.6)(2,1.6)(2,2)
\qbezier(1,2)(1,1.7)(1.5,1.7)
\qbezier(1.5,1.7)(2,1.7)(2,2)
\qbezier(1,2)(1,1.8)(1.5,1.8)
\qbezier(1.5,1.8)(2,1.8)(2,2)
\qbezier(1,2)(1,1.9)(1.5,1.9)
\qbezier(1.5,1.9)(2,1.9)(2,2)
}
\put(10,0){
\put(1.5,2){\oval[1](1,1)[t]}
\qbezier(1,2)(1,2.4)(1.5,2.4)
\qbezier(1.5,2.4)(2,2.4)(2,2)
\qbezier(1,2)(1,2.3)(1.5,2.3)
\qbezier(1.5,2.3)(2,2.3)(2,2)
\qbezier(1,2)(1,2.2)(1.5,2.2)
\qbezier(1.5,2.2)(2,2.2)(2,2)
\qbezier(1,2)(1,2.1)(1.5,2.1)
\qbezier(1.5,2.1)(2,2.1)(2,2)
\put(1.5,2){\oval[1](1,1)[b]}
\qbezier(1,2)(1,1.6)(1.5,1.6)
\qbezier(1.5,1.6)(2,1.6)(2,2)
\qbezier(1,2)(1,1.7)(1.5,1.7)
\qbezier(1.5,1.7)(2,1.7)(2,2)
\qbezier(1,2)(1,1.8)(1.5,1.8)
\qbezier(1.5,1.8)(2,1.8)(2,2)
\qbezier(1,2)(1,1.9)(1.5,1.9)
\qbezier(1.5,1.9)(2,1.9)(2,2)
}
\put(11,0){
\put(1.5,2){\oval[1](1,1)[t]}
\qbezier(1,2)(1,2.4)(1.5,2.4)
\qbezier(1.5,2.4)(2,2.4)(2,2)
\qbezier(1,2)(1,2.3)(1.5,2.3)
\qbezier(1.5,2.3)(2,2.3)(2,2)
\qbezier(1,2)(1,2.2)(1.5,2.2)
\qbezier(1.5,2.2)(2,2.2)(2,2)
\qbezier(1,2)(1,2.1)(1.5,2.1)
\qbezier(1.5,2.1)(2,2.1)(2,2)
\put(1.5,2){\oval[1](1,1)[b]}
\qbezier(1,2)(1,1.6)(1.5,1.6)
\qbezier(1.5,1.6)(2,1.6)(2,2)
\qbezier(1,2)(1,1.7)(1.5,1.7)
\qbezier(1.5,1.7)(2,1.7)(2,2)
\qbezier(1,2)(1,1.8)(1.5,1.8)
\qbezier(1.5,1.8)(2,1.8)(2,2)
\qbezier(1,2)(1,1.9)(1.5,1.9)
\qbezier(1.5,1.9)(2,1.9)(2,2)
}
\put(12,0){
\put(1.5,2){\oval[1](1,1)[t]}
\qbezier(1,2)(1,2.4)(1.5,2.4)
\qbezier(1.5,2.4)(2,2.4)(2,2)
\qbezier(1,2)(1,2.3)(1.5,2.3)
\qbezier(1.5,2.3)(2,2.3)(2,2)
\qbezier(1,2)(1,2.2)(1.5,2.2)
\qbezier(1.5,2.2)(2,2.2)(2,2)
\qbezier(1,2)(1,2.1)(1.5,2.1)
\qbezier(1.5,2.1)(2,2.1)(2,2)
\put(1.5,2){\oval[1](1,1)[b]}
\qbezier(1,2)(1,1.6)(1.5,1.6)
\qbezier(1.5,1.6)(2,1.6)(2,2)
\qbezier(1,2)(1,1.7)(1.5,1.7)
\qbezier(1.5,1.7)(2,1.7)(2,2)
\qbezier(1,2)(1,1.8)(1.5,1.8)
\qbezier(1.5,1.8)(2,1.8)(2,2)
\qbezier(1,2)(1,1.9)(1.5,1.9)
\qbezier(1.5,1.9)(2,1.9)(2,2)
}
\put(-0.2,-1){\shortstack[l]{\small{Figure 2:} \footnotesize \ Illustration of geodesics in 3D for $N=13$ and radius $r=5\ mm$ between two points A and B.}}
\end{picture}
\end{fmpage}
\end{center}
\vskip 1cm

These geodesics in three dimensions will be considered in following as light geodesics for the small scale world. It is known that one cannot predict which path is followed by photons to travel between two distant locations. If a photon is assumed to follow the geodesics (\ref{Rot}) in an homogeneous space-time as paths of least time, then the photon will have an infinity of geodesics between any distant locations $A=(0,0,0)$ and $B=R_\theta(x,\varphi_{i}(x),0)$ for $i\geq1$, and $\theta\in [0,2\pi]$ (since there exists an infinity of geodesics between two antipodal points on the sphere of center $C_{i}=((2i+1)r, 0,0)$ and radius $r$).
Moreover, since the dimension of the light wave length is between $10^{-6}\ m$ and $10^{-7}\ m$, then within $1.1\ mm$ there is 2000 times the average of the light wave length, which means that using the fluctuating geodesics defined in (\ref{Rot}) with $4r=5.5\times 10^{-7}\ m$, we obtain  4000 points of non differentiability within $1.1\ mm$, and 4000 extrema. Therefore the infinity of geodesics of least time defined by (\ref{Rot}) for $r=1.375\times 10^{-7}\ m$ appears as a straight line geodesic for the macroscopic observation.\pesp

The fluctuation of a physical system in following one of the infinity of geodesics (\ref{Rot}) for the quantum scale is preponderant if the dimension of the physical system is less than $4r$ (two successive extrema), however this motion appears to be a straight line motion if the dimension of the physical system is larger than $4r$.\pesp

The existence of an infinity of possible geodesics of least time in an homogeneous space and the existence of a big number of points of non differentiability for each geodesic induce that it is impossible to predict from which path the photon will pass through, and practically it is impossible to measure simultaneously its position and momentum with complete precision, which is consistent with the quantum indeterminacy. However knowing all possible geodesics of least time defined by (\ref{Rot}) in an homogeneous space, we will be able to understand what the photon does to travel between the narrow slit and the detector screen without determinacy, we will be able to provide a geometrical explanation of the deviation of light direction from rectilinear when light encounters edges, apertures and screens.

\section{Fluctuating Geodesics and Reflection Laws}

Let us assume that when a source emits a single photon in an homogeneous space-time, the photon travels along one path among the infinity of possible geodesics of least time defined by (\ref{Rot}) between two distant locations and illustrated in Fig.2.

\setlength{\unitlength}{.8cm}
\begin{center}
\begin{fmpage}{14.5cm}
\begin{picture}(14.5,12)
\linethickness{.1mm}
\put(5.7,7.2){$\alpha_r$}
\put(6.20,7.05){$(\!($}

\put(2,7.05){\tiny normal}

\put(6.20,6.60){$(\!($}
\put(5.7,6.5){$\alpha_i$}

  {\linethickness{.8mm}\put(7.03,4.5){\line(0,1){5}}}
  {\linethickness{.8mm}\put(15.03,4.5){\line(0,1){5}}}
{\linethickness{.3mm}\put(2.02,1){\rotatebox[origin=c]{40}{{\put(0.5,2){\vector(1,0){7}}}}}}
{\linethickness{.3mm}\put(5.84,5.06){\rotatebox[origin=c]{-40}{{\put(0.5,2){\vector(-1,0){7}}}}}}
{\linethickness{.1mm}\put(2,6.938){\line(1,0){5}}}
{\linethickness{.3mm}\put(10.02,1){\rotatebox[origin=c]{40}{{\put(0.5,2){\vector(1,0){7}}}}}}
\put(10.02,1){\rotatebox[origin=c]{40}{{{\put(0,0){
\put(1.5,2){\oval[1](1,1)[t]}
\qbezier(1,2)(1,2.4)(1.5,2.4)
\qbezier(1.5,2.4)(2,2.4)(2,2)
\qbezier(1,2)(1,2.3)(1.5,2.3)
\qbezier(1.5,2.3)(2,2.3)(2,2)
\qbezier(1,2)(1,2.2)(1.5,2.2)
\qbezier(1.5,2.2)(2,2.2)(2,2)
\qbezier(1,2)(1,2.1)(1.5,2.1)
\qbezier(1.5,2.1)(2,2.1)(2,2)
\put(1.5,2){\oval[1](1,1)[b]}
\qbezier(1,2)(1,1.6)(1.5,1.6)
\qbezier(1.5,1.6)(2,1.6)(2,2)
\qbezier(1,2)(1,1.7)(1.5,1.7)
\qbezier(1.5,1.7)(2,1.7)(2,2)
\qbezier(1,2)(1,1.8)(1.5,1.8)
\qbezier(1.5,1.8)(2,1.8)(2,2)
\qbezier(1,2)(1,1.9)(1.5,1.9)
\qbezier(1.5,1.9)(2,1.9)(2,2)
}
\put(1,0){
\put(1.5,2){\oval[1](1,1)[t]}
\qbezier(1,2)(1,2.4)(1.5,2.4)
\qbezier(1.5,2.4)(2,2.4)(2,2)
\qbezier(1,2)(1,2.3)(1.5,2.3)
\qbezier(1.5,2.3)(2,2.3)(2,2)
\qbezier(1,2)(1,2.2)(1.5,2.2)
\qbezier(1.5,2.2)(2,2.2)(2,2)
\qbezier(1,2)(1,2.1)(1.5,2.1)
\qbezier(1.5,2.1)(2,2.1)(2,2)
\put(1.5,2){\oval[1](1,1)[b]}
\qbezier(1,2)(1,1.6)(1.5,1.6)
\qbezier(1.5,1.6)(2,1.6)(2,2)
\qbezier(1,2)(1,1.7)(1.5,1.7)
\qbezier(1.5,1.7)(2,1.7)(2,2)
\qbezier(1,2)(1,1.8)(1.5,1.8)
\qbezier(1.5,1.8)(2,1.8)(2,2)
\qbezier(1,2)(1,1.9)(1.5,1.9)
\qbezier(1.5,1.9)(2,1.9)(2,2)
}
\put(2,0){
\put(1.5,2){\oval[1](1,1)[t]}
\qbezier(1,2)(1,2.4)(1.5,2.4)
\qbezier(1.5,2.4)(2,2.4)(2,2)
\qbezier(1,2)(1,2.3)(1.5,2.3)
\qbezier(1.5,2.3)(2,2.3)(2,2)
\qbezier(1,2)(1,2.2)(1.5,2.2)
\qbezier(1.5,2.2)(2,2.2)(2,2)
\qbezier(1,2)(1,2.1)(1.5,2.1)
\qbezier(1.5,2.1)(2,2.1)(2,2)
\put(1.5,2){\oval[1](1,1)[b]}
\qbezier(1,2)(1,1.6)(1.5,1.6)
\qbezier(1.5,1.6)(2,1.6)(2,2)
\qbezier(1,2)(1,1.7)(1.5,1.7)
\qbezier(1.5,1.7)(2,1.7)(2,2)
\qbezier(1,2)(1,1.8)(1.5,1.8)
\qbezier(1.5,1.8)(2,1.8)(2,2)
\qbezier(1,2)(1,1.9)(1.5,1.9)
\qbezier(1.5,1.9)(2,1.9)(2,2)
}
\put(3,0){
\put(1.5,2){\oval[1](1,1)[t]}
\qbezier(1,2)(1,2.4)(1.5,2.4)
\qbezier(1.5,2.4)(2,2.4)(2,2)
\qbezier(1,2)(1,2.3)(1.5,2.3)
\qbezier(1.5,2.3)(2,2.3)(2,2)
\qbezier(1,2)(1,2.2)(1.5,2.2)
\qbezier(1.5,2.2)(2,2.2)(2,2)
\qbezier(1,2)(1,2.1)(1.5,2.1)
\qbezier(1.5,2.1)(2,2.1)(2,2)
\put(1.5,2){\oval[1](1,1)[b]}
\qbezier(1,2)(1,1.6)(1.5,1.6)
\qbezier(1.5,1.6)(2,1.6)(2,2)
\qbezier(1,2)(1,1.7)(1.5,1.7)
\qbezier(1.5,1.7)(2,1.7)(2,2)
\qbezier(1,2)(1,1.8)(1.5,1.8)
\qbezier(1.5,1.8)(2,1.8)(2,2)
\qbezier(1,2)(1,1.9)(1.5,1.9)
\qbezier(1.5,1.9)(2,1.9)(2,2)
}
\put(4,0){
\put(1.5,2){\oval[1](1,1)[t]}
\qbezier(1,2)(1,2.4)(1.5,2.4)
\qbezier(1.5,2.4)(2,2.4)(2,2)
\qbezier(1,2)(1,2.3)(1.5,2.3)
\qbezier(1.5,2.3)(2,2.3)(2,2)
\qbezier(1,2)(1,2.2)(1.5,2.2)
\qbezier(1.5,2.2)(2,2.2)(2,2)
\qbezier(1,2)(1,2.1)(1.5,2.1)
\qbezier(1.5,2.1)(2,2.1)(2,2)
\put(1.5,2){\oval[1](1,1)[b]}
\qbezier(1,2)(1,1.6)(1.5,1.6)
\qbezier(1.5,1.6)(2,1.6)(2,2)
\qbezier(1,2)(1,1.7)(1.5,1.7)
\qbezier(1.5,1.7)(2,1.7)(2,2)
\qbezier(1,2)(1,1.8)(1.5,1.8)
\qbezier(1.5,1.8)(2,1.8)(2,2)
\qbezier(1,2)(1,1.9)(1.5,1.9)
\qbezier(1.5,1.9)(2,1.9)(2,2)
}
\put(5,0){
\put(1.5,2){\oval[1](1,1)[t]}
\qbezier(1,2)(1,2.4)(1.5,2.4)
\qbezier(1.5,2.4)(2,2.4)(2,2)
\qbezier(1,2)(1,2.3)(1.5,2.3)
\qbezier(1.5,2.3)(2,2.3)(2,2)
\qbezier(1,2)(1,2.2)(1.5,2.2)
\qbezier(1.5,2.2)(2,2.2)(2,2)
\qbezier(1,2)(1,2.1)(1.5,2.1)
\qbezier(1.5,2.1)(2,2.1)(2,2)
\put(1.5,2){\oval[1](1,1)[b]}
\qbezier(1,2)(1,1.6)(1.5,1.6)
\qbezier(1.5,1.6)(2,1.6)(2,2)
\qbezier(1,2)(1,1.7)(1.5,1.7)
\qbezier(1.5,1.7)(2,1.7)(2,2)
\qbezier(1,2)(1,1.8)(1.5,1.8)
\qbezier(1.5,1.8)(2,1.8)(2,2)
\qbezier(1,2)(1,1.9)(1.5,1.9)
\qbezier(1.5,1.9)(2,1.9)(2,2)
}}}}}
{\linethickness{.3mm}\put(13.84,5.06){\rotatebox[origin=c]{-40}{{\put(0.5,2){\vector(-1,0){7}}}}}}
{\linethickness{.1mm}\put(10,6.938){\line(1,0){5}}}
\put(8.73,9.35){\rotatebox[origin=c]{-40}{{{\put(0,0){
\put(1.5,2){\oval[1](1,1)[t]}
\qbezier(1,2)(1,2.4)(1.5,2.4)
\qbezier(1.5,2.4)(2,2.4)(2,2)
\qbezier(1,2)(1,2.3)(1.5,2.3)
\qbezier(1.5,2.3)(2,2.3)(2,2)
\qbezier(1,2)(1,2.2)(1.5,2.2)
\qbezier(1.5,2.2)(2,2.2)(2,2)
\qbezier(1,2)(1,2.1)(1.5,2.1)
\qbezier(1.5,2.1)(2,2.1)(2,2)
\put(1.5,2){\oval[1](1,1)[b]}
\qbezier(1,2)(1,1.6)(1.5,1.6)
\qbezier(1.5,1.6)(2,1.6)(2,2)
\qbezier(1,2)(1,1.7)(1.5,1.7)
\qbezier(1.5,1.7)(2,1.7)(2,2)
\qbezier(1,2)(1,1.8)(1.5,1.8)
\qbezier(1.5,1.8)(2,1.8)(2,2)
\qbezier(1,2)(1,1.9)(1.5,1.9)
\qbezier(1.5,1.9)(2,1.9)(2,2)
}
\put(1,0){
\put(1.5,2){\oval[1](1,1)[t]}
\qbezier(1,2)(1,2.4)(1.5,2.4)
\qbezier(1.5,2.4)(2,2.4)(2,2)
\qbezier(1,2)(1,2.3)(1.5,2.3)
\qbezier(1.5,2.3)(2,2.3)(2,2)
\qbezier(1,2)(1,2.2)(1.5,2.2)
\qbezier(1.5,2.2)(2,2.2)(2,2)
\qbezier(1,2)(1,2.1)(1.5,2.1)
\qbezier(1.5,2.1)(2,2.1)(2,2)
\put(1.5,2){\oval[1](1,1)[b]}
\qbezier(1,2)(1,1.6)(1.5,1.6)
\qbezier(1.5,1.6)(2,1.6)(2,2)
\qbezier(1,2)(1,1.7)(1.5,1.7)
\qbezier(1.5,1.7)(2,1.7)(2,2)
\qbezier(1,2)(1,1.8)(1.5,1.8)
\qbezier(1.5,1.8)(2,1.8)(2,2)
\qbezier(1,2)(1,1.9)(1.5,1.9)
\qbezier(1.5,1.9)(2,1.9)(2,2)
}
\put(2,0){
\put(1.5,2){\oval[1](1,1)[t]}
\qbezier(1,2)(1,2.4)(1.5,2.4)
\qbezier(1.5,2.4)(2,2.4)(2,2)
\qbezier(1,2)(1,2.3)(1.5,2.3)
\qbezier(1.5,2.3)(2,2.3)(2,2)
\qbezier(1,2)(1,2.2)(1.5,2.2)
\qbezier(1.5,2.2)(2,2.2)(2,2)
\qbezier(1,2)(1,2.1)(1.5,2.1)
\qbezier(1.5,2.1)(2,2.1)(2,2)
\put(1.5,2){\oval[1](1,1)[b]}
\qbezier(1,2)(1,1.6)(1.5,1.6)
\qbezier(1.5,1.6)(2,1.6)(2,2)
\qbezier(1,2)(1,1.7)(1.5,1.7)
\qbezier(1.5,1.7)(2,1.7)(2,2)
\qbezier(1,2)(1,1.8)(1.5,1.8)
\qbezier(1.5,1.8)(2,1.8)(2,2)
\qbezier(1,2)(1,1.9)(1.5,1.9)
\qbezier(1.5,1.9)(2,1.9)(2,2)
}
\put(3,0){
\put(1.5,2){\oval[1](1,1)[t]}
\qbezier(1,2)(1,2.4)(1.5,2.4)
\qbezier(1.5,2.4)(2,2.4)(2,2)
\qbezier(1,2)(1,2.3)(1.5,2.3)
\qbezier(1.5,2.3)(2,2.3)(2,2)
\qbezier(1,2)(1,2.2)(1.5,2.2)
\qbezier(1.5,2.2)(2,2.2)(2,2)
\qbezier(1,2)(1,2.1)(1.5,2.1)
\qbezier(1.5,2.1)(2,2.1)(2,2)
\put(1.5,2){\oval[1](1,1)[b]}
\qbezier(1,2)(1,1.6)(1.5,1.6)
\qbezier(1.5,1.6)(2,1.6)(2,2)
\qbezier(1,2)(1,1.7)(1.5,1.7)
\qbezier(1.5,1.7)(2,1.7)(2,2)
\qbezier(1,2)(1,1.8)(1.5,1.8)
\qbezier(1.5,1.8)(2,1.8)(2,2)
\qbezier(1,2)(1,1.9)(1.5,1.9)
\qbezier(1.5,1.9)(2,1.9)(2,2)
}
\put(4,0){
\put(1.5,2){\oval[1](1,1)[t]}
\qbezier(1,2)(1,2.4)(1.5,2.4)
\qbezier(1.5,2.4)(2,2.4)(2,2)
\qbezier(1,2)(1,2.3)(1.5,2.3)
\qbezier(1.5,2.3)(2,2.3)(2,2)
\qbezier(1,2)(1,2.2)(1.5,2.2)
\qbezier(1.5,2.2)(2,2.2)(2,2)
\qbezier(1,2)(1,2.1)(1.5,2.1)
\qbezier(1.5,2.1)(2,2.1)(2,2)
\put(1.5,2){\oval[1](1,1)[b]}
\qbezier(1,2)(1,1.6)(1.5,1.6)
\qbezier(1.5,1.6)(2,1.6)(2,2)
\qbezier(1,2)(1,1.7)(1.5,1.7)
\qbezier(1.5,1.7)(2,1.7)(2,2)
\qbezier(1,2)(1,1.8)(1.5,1.8)
\qbezier(1.5,1.8)(2,1.8)(2,2)
\qbezier(1,2)(1,1.9)(1.5,1.9)
\qbezier(1.5,1.9)(2,1.9)(2,2)
}
\put(5,0){
\put(1.5,2){\oval[1](1,1)[t]}
\qbezier(1,2)(1,2.4)(1.5,2.4)
\qbezier(1.5,2.4)(2,2.4)(2,2)
\qbezier(1,2)(1,2.3)(1.5,2.3)
\qbezier(1.5,2.3)(2,2.3)(2,2)
\qbezier(1,2)(1,2.2)(1.5,2.2)
\qbezier(1.5,2.2)(2,2.2)(2,2)
\qbezier(1,2)(1,2.1)(1.5,2.1)
\qbezier(1.5,2.1)(2,2.1)(2,2)
\put(1.5,2){\oval[1](1,1)[b]}
\qbezier(1,2)(1,1.6)(1.5,1.6)
\qbezier(1.5,1.6)(2,1.6)(2,2)
\qbezier(1,2)(1,1.7)(1.5,1.7)
\qbezier(1.5,1.7)(2,1.7)(2,2)
\qbezier(1,2)(1,1.8)(1.5,1.8)
\qbezier(1.5,1.8)(2,1.8)(2,2)
\qbezier(1,2)(1,1.9)(1.5,1.9)
\qbezier(1.5,1.9)(2,1.9)(2,2)
}}}}}
\put(9.5,2.5){\footnotesize Incident Geodesics}
\put(9.5,11.5){\footnotesize Reflected Geodesics}
\put(1.5,2.5){\footnotesize Incident Geodesics Direction}
\put(1.5,11.5){\footnotesize Reflected Geodesics Direction}
\put(6.5,4){\footnotesize Interface}
\put(14.5,4){\footnotesize Interface}
\put(6,1){$(a)$}
\put(12,1){$(b)$}
\put(0,-2.5){\shortstack[l]{\small{Figure 3:} \footnotesize Fluctuating geodesics reflection based on symmetric with respect to the normal
 line to\\ \footnotesize the interface that passes through the intersection point of incident geodesics axis and
reflected\\ \footnotesize geodesics axis. (a) Illustration of reflection of the geodesics axis. (b) illustration of reflection of all\\ \footnotesize geodesics following the geodesics axis.}}
\end{picture}
\end{fmpage}
\end{center}
\vskip2cm
The $x$-axis represents the {\it geodesics axis} that determines the light overall direction, and the $yz$-plane represents the plane of geodesics oscillation. The geodesics given by the graph (\ref{Rot}) in an homogeneous space-time are resultant of composition of oscillations in the $yz$-plane and a translation with increasing-decreasing magnitude directed by the {\it geodesics axis}. The oscillations determine the period, the frequency and the amplitude of the geodesics.\pesp

Let us assume in the following that when a photon follows one fluctuating geodesic from the infinity of paths of least time defined by (\ref{Rot}) and encounters an interface that induces a specular reflection (mirror for example), the geodesics axis of the photon's path verifies the classical law of reflection, and the angle which the incident {\it geodesics axis} makes with the normal is equal to the angle which the reflected {\it geodesics axis} makes with the normal (see Fig.3, (a) and (b)). The local mechanism of reflection of the fluctuating geodesics (\ref{Rot}) with different form of interface will be the subject of a further work.\esp

\setlength{\unitlength}{.7cm}
\begin{center}
\begin{fmpage}{9cm}
\begin{picture}(14,9)
\linethickness{.1mm}
{\linethickness{.1mm}\put(0.73,6){\rotatebox[origin=c]{-40}{{\put(0.5,2){\vector(1,0){6}}}}}}
\put(0.73,6){\rotatebox[origin=c]{-40}{{{\put(0,0){
\put(1.5,2){\oval[2](2,2)[t]}
\put(1.5,2){\oval[2](2,2)[b]}
}
\put(2,0){
\put(1.5,2){\oval[2](2,2)[t]}
\put(1.5,2){\oval[2](2,2)[b]}
}
\put(4,0){
\put(1.5,2){\oval[2](2,2)[t]}
\put(1.5,2){{\rotatebox[origin=c]{80}{\oval[2](2,2)[t]}}}
}
}}}}
\put(5.46,3.6){\rotatebox[origin=c]{141}{{\vector(1,0){.5}}}}
\put(5.46,3.6){\rotatebox[origin=c]{42.8}{{\vector(1,0){.5}}}}
\bezier{10}(5.45,3.6)(5.45,3.8)(5.45,4)
\put(7,3.6){\rotatebox[origin=c]{141}{{\vector(1,0){.5}}}}
\put(7,3.6){\rotatebox[origin=c]{42.8}{{\vector(1,0){.5}}}}

\bezier{10}(6.99,3.6)(6.99,3.8)(6.99,4)

\qbezier(5.44,3.58)(6.215,4.28)(6.99,3.58)

{\linethickness{.1mm}\put(5.05,1.51){\rotatebox[origin=c]{40}{{\put(0.5,2){\vector(1,0){7}}}}}}
\put(5.084,1.51){\rotatebox[origin=c]{40}{{{\put(0,0){
\put(1.5,2){\oval[2](2,2)[t]}
\put(1.5,2){{\rotatebox[origin=c]{-80}{\oval[2](2,2)[t]}}}
}
\put(2,0){
\put(1.5,2){\oval[2](2,2)[t]}
\put(1.5,2){\oval[2](2,2)[b]}
}
\put(4,0){
\put(1.5,2){\oval[2](2,2)[t]}
\put(1.5,2){\oval[2](2,2)[b]}
}
}}}}

{\linethickness{.3mm}\put(2,3.60){\line(1,0){10}}}
{\linethickness{.3mm}\bezier{40}(6.21,3.61)(6.21,5.6)(6.21,8)}
\put(5.3,5.1){\footnotesize A}
\put(5.3,3){\footnotesize B}
\put(6.9,3){\footnotesize C}
\put(7,5.1){\footnotesize D}
\put(1.3,8.5){\footnotesize Incident Geodesics}
\put(8,8.5){\footnotesize Reflected Geodesics}
\put(9.9,3){\footnotesize Interface}
\put(-4.5,-2){\shortstack[l]{\small{Figure 4:} \footnotesize Illustration of reflection of fluctuating geodesics in 2D, where the geodesics axis intersection\\ \footnotesize (of of the incident and reflected geodesics) is not in the interface. The incident and reflected geodesics\\ \footnotesize are symmetric with respect to the interface normal line.}}
\end{picture}
\end{fmpage}
\end{center}
\vskip1.5cm
Nevertheless the illustration of the reflection in two dimensions given in Fig.4 provides an explanation of the local reflection that leads to the global reflection of the geodesics axis: if the photon follows the upper geodesic near the antipodal point $A$, then the photon follows the arc $\stackrel\frown{AB}$,  and is locally reflected at the incident point $B$ following the law of reflection (the tangent line of the photon's incident path and the tangent line of the photon's reflected path at the point of incidence are symmetric with respect to the normal line that passes through the point of incidence). Afterward, the photon follows the arc $\stackrel\frown{BC}$, and is reflected a second time at the incident point $C$ (following the law of reflection), then it follows the arc $\stackrel\frown{CD}$ to reach the antipodal point $D$ and to travel following the upper geodesic. If the photon follows the lower geodesic near the antipodal point $A$, then the photon follows the arc $\stackrel\frown{AC}$, and is locally reflected at the incident point $C$ following the law of reflection. Afterward, the photon follows the arc $\stackrel\frown{CB}$, and is reflected a second time at the incident point $B$ (following the law of reflection), then it follows the arc $\stackrel\frown{BD}$ to reach the antipodal point $D$ and to travel following the lower geodesic in the direction of the reflected geodesics axis.

\section{Fluctuation and Diffraction of Light}

Let us consider a photon that travels along one path among the infinity of possible geodesics of least time defined by (\ref{Rot}) between two distant locations in a given direction and encounters a barrier that has an opening. The reflection of the photon on the barrier side depends on the path of least time chosen from the infinity of possible geodesics defined by (\ref{Rot}) (on the local fluctuation) and the position of the geodesics axis. Three possible cases can occur when the light geodesics encounter the corner of the obstacle of width $l$ (see Fig.5):\pesp

i) the geodesics axis is parallel and above the side of width $l$ (the horizontal side of the barrier illustrated in (a), Fig.5); \pesp

ii) the geodesics axis is parallel and on the side of width $l$ (the horizontal side of the barrier illustrated in (b), Fig.5);\pesp

iii) the geodesics axis is parallel and below the side of width $l$ (the horizontal side of the barrier illustrated in (c), Fig.5).\esp

\setlength{\unitlength}{.7mm}
\begin{center}
\begin{fmpage}{14.5cm}
\begin{picture}(100,50)

{\linethickness{0.3mm}\multiput(10,32)(0,20){1}{\line(1,0){10}}
  \multiput(10,12)(10,0){2}{\line(0,1){20}}}

{\linethickness{0.4mm}  \put(-0.4,35){\vector(1,0){35}}}
\put(14,14){$l$}
  \put(10,11){\vector(1,0){10}}
  \put(20,11){\vector(-1,0){10}}


{\linethickness{0.3mm}\multiput(90,32)(0,20){1}{\line(1,0){10}}
\multiput(90,12)(10,0){2}{\line(0,1){20}}}
{\linethickness{0.4mm}\put(79.7,32){\vector(1,0){35}}}

  \put(94,14){$l$}
  \put(90,11){\vector(1,0){10}}
  \put(100,11){\vector(-1,0){10}}


  {\linethickness{0.3mm}\multiput(171,32)(0,20){1}{\line(1,0){10}}
  \multiput(171,12)(10,0){2}{\line(0,1){20}}}
{\linethickness{0.4mm}\put(135.5,29){\vector(2,0){35}}}

\put(175,14){$l$}
  \put(171,11){\vector(1,0){10}}
  \put(181,11){\vector(-1,0){10}}

\put(11,2){(a)}
\put(91,2){(b)}
\put(172,2){(c)}
\put(3,-10){\shortstack[l]{\small{Figure 5:} \footnotesize Illustration of the geodesics axis position with respect to the barrier of width $l$.}}
\end{picture}
\end{fmpage}
\end{center}
\vskip1cm

 For the cases (i) and (ii), using the reflection law locally, the photon passes through the opening and conserves the direction of its geodesics axis, meanwhile for the case (iii), using the reflection law locally and an additional assumption, the photon passes through the opening and changes the direction of its geodesics axis, which creates the light diffraction. Indeed:

\setlength{\unitlength}{.7mm}
\begin{center}
\begin{fmpage}{14.5cm}
\begin{picture}(100,60)

{\linethickness{0.3mm}\multiput(10,32)(0,20){1}{\line(1,0){10}}
  \multiput(10,12)(10,0){2}{\line(0,1){20}}}

  \put(3.9,35){\circle{9}}
  \put(12.9,35){\oval[9](9,9)[t]}
   \put(12.9,35){\rotatebox[origin=c]{45}{{\oval[9](9,9)[t]}}}
 \qbezier(17.4,35)(17.4,33)(18.4,32)
 \qbezier(18.4,32)(19.4,33)(19.4,35)
  \put(23.9,35){\oval[9](9,9)[t]}
  \put(32.9,35){\oval[9](9,9)[b]}
  \put(41.9,35){\oval[9](9,9)[t]}
  \put(50.9,35){\oval[9](9,9)[b]}
  {\linethickness{0.4mm}\put(-0.4,35){\vector(1,0){60}}}
  {\linethickness{.1mm}\put(18.4,32){\vector(-0.8,1){5}}}
  {\linethickness{.1mm}\put(18.4,32){\vector(0.8,1){5}}}

  {\linethickness{.1mm}\put(18.4,32){\line(0,1){10}}}
  {\linethickness{.1mm}\put(17,45){n}}
{\linethickness{.1mm}\put(8,41){$\overrightarrow{T_i}$}}
{\linethickness{.1mm}\put(24,41){$\overrightarrow{T_r}$}}

  {\linethickness{0.3mm}\multiput(90,32)(0,20){1}{\line(1,0){10}}
  \multiput(90,12)(10,0){2}{\line(0,1){20}}}
  \put(84,32){\circle{9}}
  \put(93,32){\oval[9](9,9)[t]}
   \put(93,32){\rotatebox[origin=c]{45}{{\oval[9](9,9)[t]}}}
  \put(102,32){\oval[9](9,9)[t]}
  \put(110.9,32){\oval[9](9,9)[b]}
  \put(119.9,32){\oval[9](9,9)[t]}
  \put(128.9,32){\oval[9](9,9)[b]}

{\linethickness{0.4mm} \put(79.7,32){\vector(1,0){58}}}
  {\linethickness{.1mm}\put(97.4,32){\vector(0,1){7}}}

  \put(95,41){$\overrightarrow{T}$}


  {\linethickness{0.3mm}\multiput(161,32)(0,20){1}{\line(1,0){10}}
  \multiput(161,12)(10,0){2}{\line(0,1){20}}}
  \put(155,29){\circle{9}}
  \put(164,29){\rotatebox[origin=c]{45}{{\oval[9](9,9)[t]}}}
{\linethickness{.4mm}\put(167.4,32){\vector(2,-2){28}}}
{\linethickness{0.4mm} \put(150.5,29){\vector(2,0){11}}}
{\linethickness{.1mm}\put(167.4,32){\vector(-3,3){7}}}
{\linethickness{.1mm}\put(167.4,32){\vector(1,1){7}}}
{\linethickness{.1mm}\put(167.4,32){\line(0,1){7}}}
{\linethickness{.1mm}\put(166,43){n}}
{\linethickness{.1mm}\put(155,41){$\overrightarrow{T_i}$}}
{\linethickness{.1mm}\put(175,41){$\overrightarrow{T_r}$}}

\put(170.7,29){\rotatebox[origin=c]{-45}{{\oval[9](9,9)[t]}}}
\put(177,22.6){\rotatebox[origin=c]{-45}{{\oval[9](9,9)[b]}}}
\put(183.2,16.1){\rotatebox[origin=c]{-45}{{\oval[9](9,9)[t]}}}
\put(189.7,9.9){\rotatebox[origin=c]{-45}{{\oval[9](9,9)[b]}}}
\put(14,25){$P$}
\put(94,25){$P$}
\put(165,25){$P$}
\put(11,2){(a)}
\put(91,2){(b)}
\put(162,2){(c)}
\put(0,-10){\shortstack[l]{\small{Figure 6:} \footnotesize \ Illustration of one geodesics deviation with respect to the barrier of width $l$.}}
\end{picture}
\end{fmpage}
\end{center}
\vskip 1cm

i) for the case (a) Fig.5, the lower part of the infinity of geodesics (\ref{Rot}), that encounters the vertical interface orthogonal to the geodesics axis, is obstructed (reflected by the vertical side of the barrier) and the photon near the corner has only the possibility to travel with the upper geodesics and the part of geodesics that does not encounter the vertical interface. If the photon follows one of the geodesics that do not cross the vertical interface, and encounters the horizontal side of the barrier of width $l$ at the point of incidence $P$ (see (a), for one geodesic Fig.6), it will be locally reflected following the laws of reflection (the tangent line with unit vector $\overrightarrow{T_i}$ of the photon's incident path and the tangent line with the unit vector $\overrightarrow{T_r}$ of the photon's reflected path at the point of incidence $P$ are symmetric with respect to the normal line $n$ that passes through the point of incidence $P$). When the photon is reflected, following the chosen path, it crosses the incident geodesics axis and undergoes a displacement that follows it. The local reflection (on the horizontal side of the barrier) changes only the antipodal points and conserves the overall displacement direction determined by the geodesics axis (only the upper part of the  geodesics (\ref{Rot}) and a part of the lower geodesics of (\ref{Rot}) that does not encounter the vertical barrier side pass through without any change in the incident geodesics axis).\pesp

ii) for the case (b) Fig.5, the lower part of the infinity of geodesics (\ref{Rot}), that encounters the vertical interface, is obstructed (reflected by the vertical side of the barrier) and the photon near the corner has only the possibility to travel with the upper geodesics. When the photon, by following the geodesics, encounters the horizontal side of the barrier of width $l$ (see (b), Fig.6) at the point of incidence $P$, it will be locally reflected following the laws of reflection (the tangent line of unit vector $\overrightarrow{T_i}$ of the photon's incident path is vertical as well as the tangent line of unit vector $\overrightarrow{T_r}$ of the photon's reflected path at the point of incidence $P$) and undergoes a displacement that follows the incident geodesics axis. The local reflection on the horizontal barrier side of width $l$ does not change the antipodal points of the geodesics and conserves the overall displacement direction determined by the geodesics axis (only the upper part of the geodesics (\ref{Rot}) passes through without any change in the incident geodesics axis).\pesp

iii) for the case (c) Fig.5, the lower part of the infinity of geodesics (\ref{Rot}), that encounters the vertical interface, is obstructed (reflected by the vertical side of the barrier) as well as a part of the upper geodesics. The photon near the corner has only the possibility to travel with the upper geodesics that do not encounter the barrier. When the photon, by following the geodesics, encounters the horizontal side of the barrier of width $l$ (see (c), Fig.6) at the point of incidence $P$, it will be locally reflected following the laws of reflection (the tangent line of unit vector $\overrightarrow{T_i}$ of the photon's incident path and the tangent line of the unit vector $\overrightarrow{T_r}$ of the photon's reflected path at the point of incidence $P$ are symmetric with respect to the normal line that passes through the point of incidence $P$) and undergoes a displacement that follows a constructed geodesics axis (by considering the point of incidence $P$ as a starting antipodal point for a new geodesic direction since it does not encounter neither obstacle nor the incident geodesics axis) perpendicular to the tangent vector $\overrightarrow{T_r}$ on the point of incidence $P$. The local reflection on the horizontal barrier side of width $l$ changes the antipodal points of the geodesics and deviates the overall displacement direction by constructing a new geodesics axis that is perpendicular to $\overrightarrow{T_r}$ (only the upper part of the geodesics that do not encounter the vertical barrier interface passes through with deviation of the geodesics axis). The deviation of the geodesics axis from horizontal is related to the angle of the tangent vector $\overrightarrow{T_r}$ with respect to the normal line at the point of incidence $P$. Indeed, the farther the incident geodesics axis is from the obstacle corner (down the corner in Fig.7, (a)), the bigger the angle is between the normal line and the tangent vector $\overrightarrow{T_r}$, and the more the new geodesics axis deviates from the horizontal beyond the barrier (such that it remains orthogonal to the reflected tangent vector $\overrightarrow{T_r}$ at the point of incidence $P$). The bigger the obliquity of the tangent $\overrightarrow{T_r}$ at the point of incidence $P$ is, the bigger the deviation of the geodesics axis is (see illustration in 2D of geodesics (a), Fig.7).
\pesp
The deviation of the overall direction (the geodesics axis) of the photon's geodesic beyond the boundary, when the incident geodesics axis is totally obstructed by the vertical side of the barrier meanwhile the photon's geodesic bypasses the vertical side of the barrier and is reflected by the horizontal side of the barrier, is obtained under the assumption that the point of incidence $P$, in the horizontal side of the barrier of width $L$, becomes antipodal point for the new geodesic.

\setlength{\unitlength}{1mm}
\begin{center}
\begin{fmpage}{14.5cm}

\end{fmpage}
\end{center}
\vskip2cm
The photon cannot undergo in the direction of the incident geodesics axis since this axis is totally obstructed by the vertical side of the barrier. The geodesics axis represents the overall direction of the photon's geodesics, which is a resultant of a composition of oscillations in the $yz$-plane and a translation with variable magnitude in the $x$-axis. The obstruction of the translation in the $x$-direction induces the deviation of the new photon's geodesic beyond the barrier. The farther the geodesics axis (of the infinity of geodesics given by (\ref{Rot})) is down the barrier's corner, the bigger the geodesics axis deviation is beyond the barrier, the smaller the number of geodesics is, that bypass the vertical side of the barrier and deviate in a new geodesics axis direction. This induces that the light intensity at any point on a given detector screen beyond the slit screen depends on the number of geodesics that bypass the barriers and are reflected by the barrier horizontal side.\pesp

The diffraction of the light geodesics is obtained as the photons pass following a part of fluctuating geodesics around the edge of the barrier. The infinity of geodesics reflected by the barrier side of width $l$ creates a new geodesics axis (see illustration of one reflected geodesic (b) Fig.7) and deviates into the region beyond the barrier with an increasing angle until all the light geodesics are totally obstructed by the vertical side of the barrier, which explains the deviation of the light from rectilinear direction.\pesp
If photons, by traveling via the infinity of geodesics defined by (\ref{Rot}), encounter a barrier that has an opening of dimension $d=2r$, the part of the geodesics that is not reflected by the vertical edge of the barrier passes through the opening, flares out (diffracts) into the region beyond the barrier, and the flaring is due to the reflection of the photons on the horizontal barrier's edge (see Fig.8). The superimposition of two distant copies, of the diffraction illustrated in (Fig.8), on the same graph, reproduces the interference pattern of fringes visible in the whole intersection region between geodesics, similar to the interference observed in the detector screen of Young's double-slit experiment (see \cite{BF}).

\setlength{\unitlength}{1mm}
\begin{center}
\begin{fmpage}{14.5cm}

\end{fmpage}
\end{center}
\vskip2cm

\section{Example: Laser Beam Diffraction}

A laser beam that passes through a vertical slit gives a predicted laser spot on a distant detector screen. If one makes the vertical slit narrower, the spot in the detector screen will get narrower together with the slit's width as predicted, until a certain small width. Then the laser spot on the distant detector screen becomes horizontally wider and wider as the slit continues to be narrower and narrower, which is an extremely non intuitive behavior. This non intuitive behavior finds its interpretation in the Heisenberg's uncertainty relation. Nevertheless it can be understood with a conclusive simple geometrical interpretation using the infinity of geodesics defined by (\ref{Rot}) and represented in Fig.2.

\subsection{Mechanism of the Geometrical Interpretation}

Consider for more simplicity a polarization of the geodesics (\ref{Rot}) in a given plane, and consider a given vertical slit in the same plane with width $w_0=4r$. The polarization of the infinity of geodesics (\ref{Rot}) in the slit's plane gives too possible opposite fluctuating geodesics with the same geodesics axis (the black and the grey geodesics in Fig.\ref{Geod}). The existence of a big number of points of non differentiability at the average scale of light wave length (for $4r=5.5\times 10^{-7}\ m$) induces that it is impossible to predict from which path the photon will pass through since each of the two fluctuating geodesics represents the path of least time in an homogeneous space.

The antipodal points of the polarized geodesics (\ref{Rot}) are points of non differentiability, therefore within $1.1mm$ there are 4000 antipodal points at the average scale of light wave length (see subsection \ref{Fluct}). Moreover at each antipodal point the photon has two possibilities to change path in the polarization plane (that is to say, using formula (\ref{least}), any photon that follows paths of least time using the polarized geodesics (\ref{Rot}) in a given plane has $2^{4000}$ possible paths within $1.1mm$), which makes the experimental prediction at this scale practically impossible, and consistent with the quantum indeterminacy.

\begin{figure}[h!]
\begin{minipage}[t]{7cm}
\centering
\fbox{\includegraphics[width=5cm]{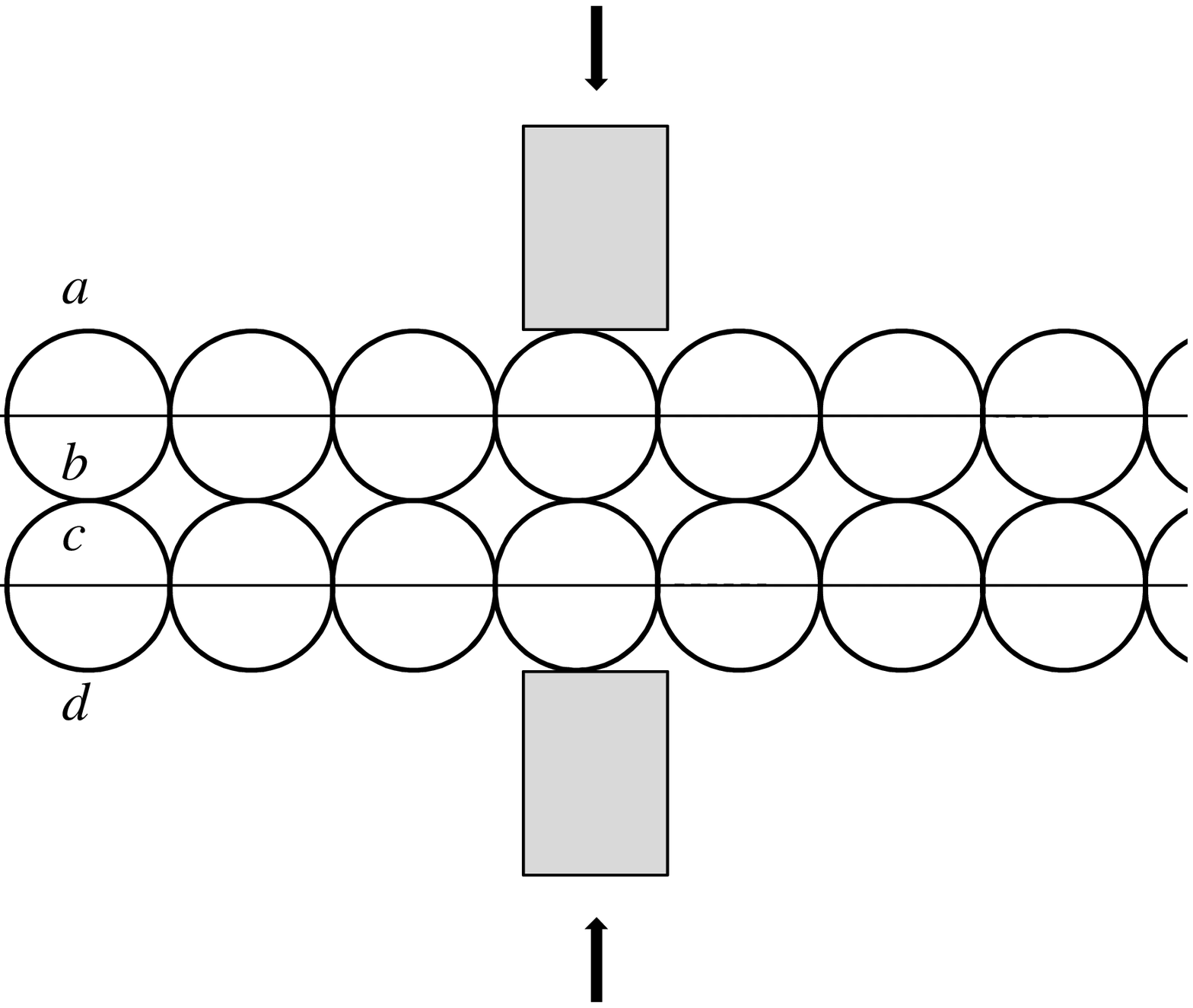}}
\caption{\footnotesize Illustration of  4 fluctuating geodesics $a$, $b$, $c$ and $d$, that pass from the left to the right through a single slit of width $w_0=4r$. The geodesics $a$ and $b$ represent a polarization of the geodesics (\ref{Rot}) in the same plane as the slit. The geodesics $a$ and $b$ (respectively $c$ and $d$ ) represent two fluctuating geodesics in anti-phase with the same geodesics axis. $a$ is the upper geodesic and $b$ the down geodesic. In this illustration $r=3.25mm$.}\label{Fig01}
\end{minipage}
\hspace*{\fill}
\begin{minipage}[t]{7cm}
\centering
\fbox{\includegraphics[width=5cm]{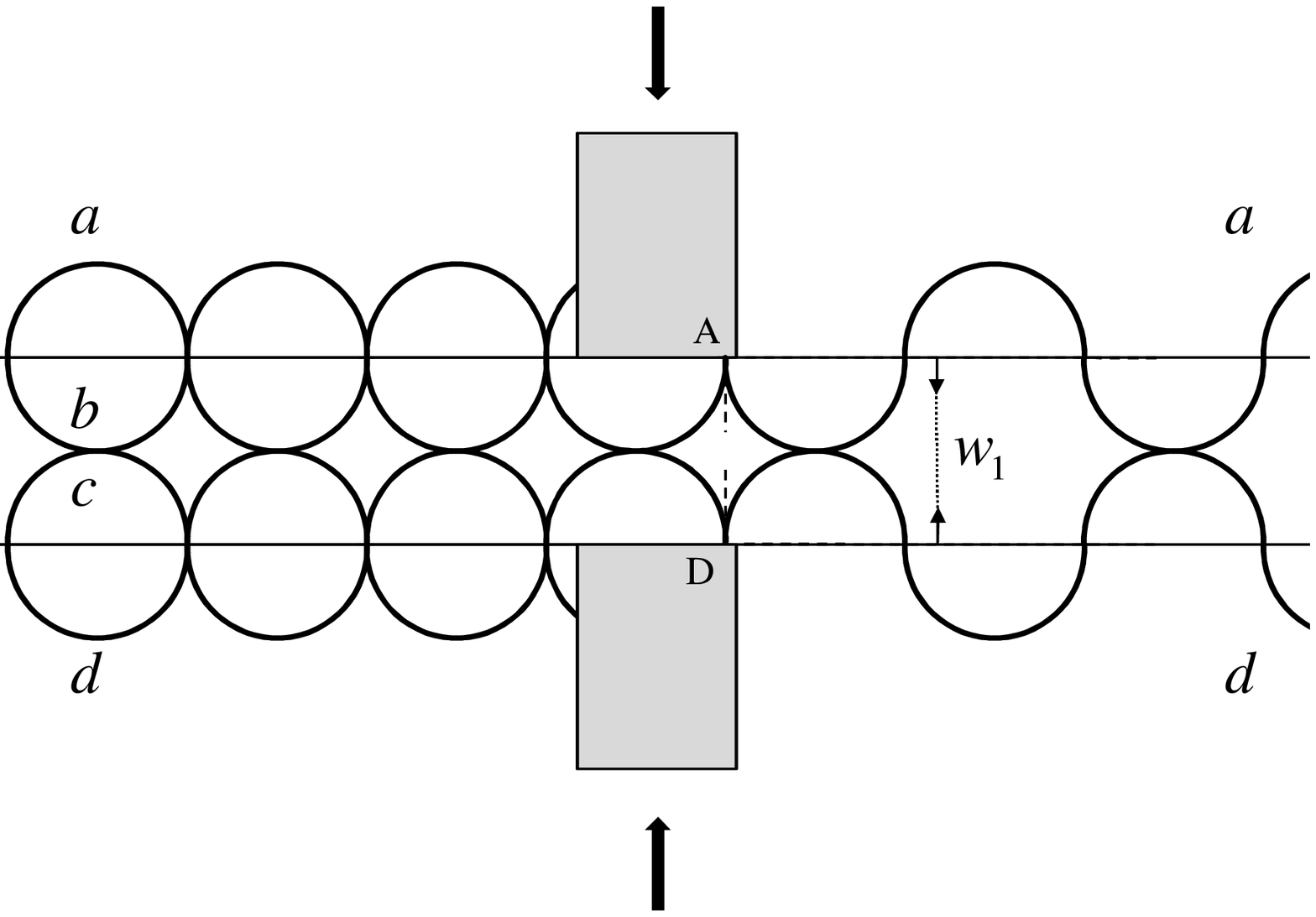}}
\caption{\footnotesize If the width of the slit $w_1$ is equal to $2r$, the lower geodesic $b$ and the upper geodesic $c$ are obstructed by the vertical sides of the slit. The upper geodesic $a$ and the lower geodesic $d$ are reflected by the horizontal side of the slit and pass through the slit to form a reduced spot in a distant detector screen.}\label{Fig02}
\end{minipage}
\end{figure}

Nevertheless to understand how the horizontal side of the slit creates the diffraction of a laser beam, it is convenient to observe what happens to a reduced number of geodesics in a given plane as the slit becomes narrower and narrower to figure out the equation that generates the deviation (diffraction) of light geodesics. In this purpose, consider four possible geodesics denoted by $a$, $b$, $c$ and $d$ that pass through the vertical slit of width $4r$ with parallel geodesics axis as illustrated in Fig.\ref{Fig01}, where the geodesics $a$ and $b$, respectively $c$ and $d$,  have the same geodesics axis (the geodesics $a$ and $b$, respectively $c$ and $d$, are obtained after polarization of the geodesics (\ref{Rot}) in the same plane).

\pesp
\pesp
\begin{figure}[h!]
\begin{minipage}[t]{7cm}
\centering
\fbox{\includegraphics[width=5cm]{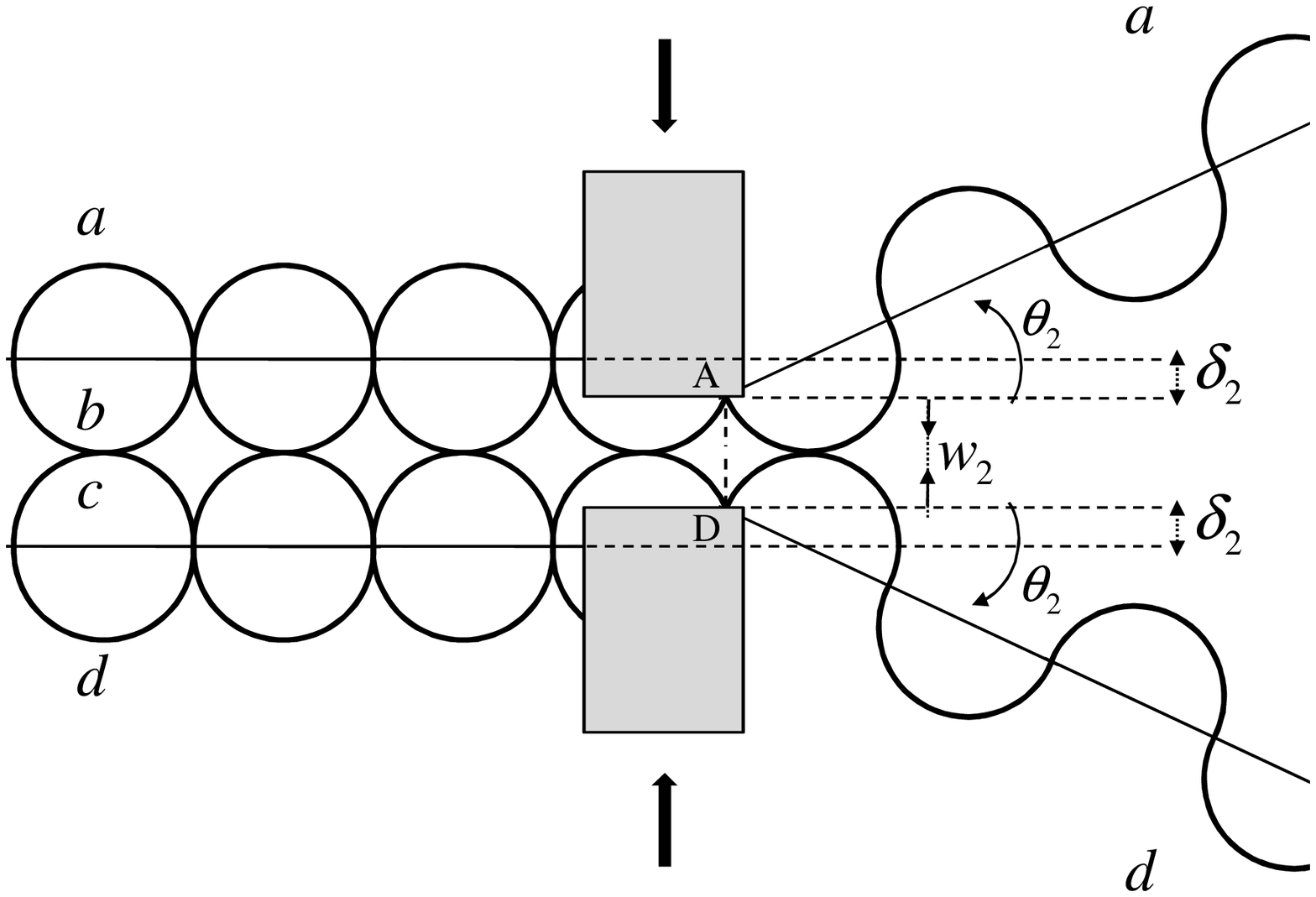}}
\caption{\footnotesize The deviation of the geodesics axis satisfies $\sin\theta_2={\delta_2\over r}$, when $2\delta_2+w_2=2r$ .}\label{Fig03}
\end{minipage}
\hspace*{\fill}
\begin{minipage}[t]{7cm}
\centering
\fbox{\includegraphics[width=5cm]{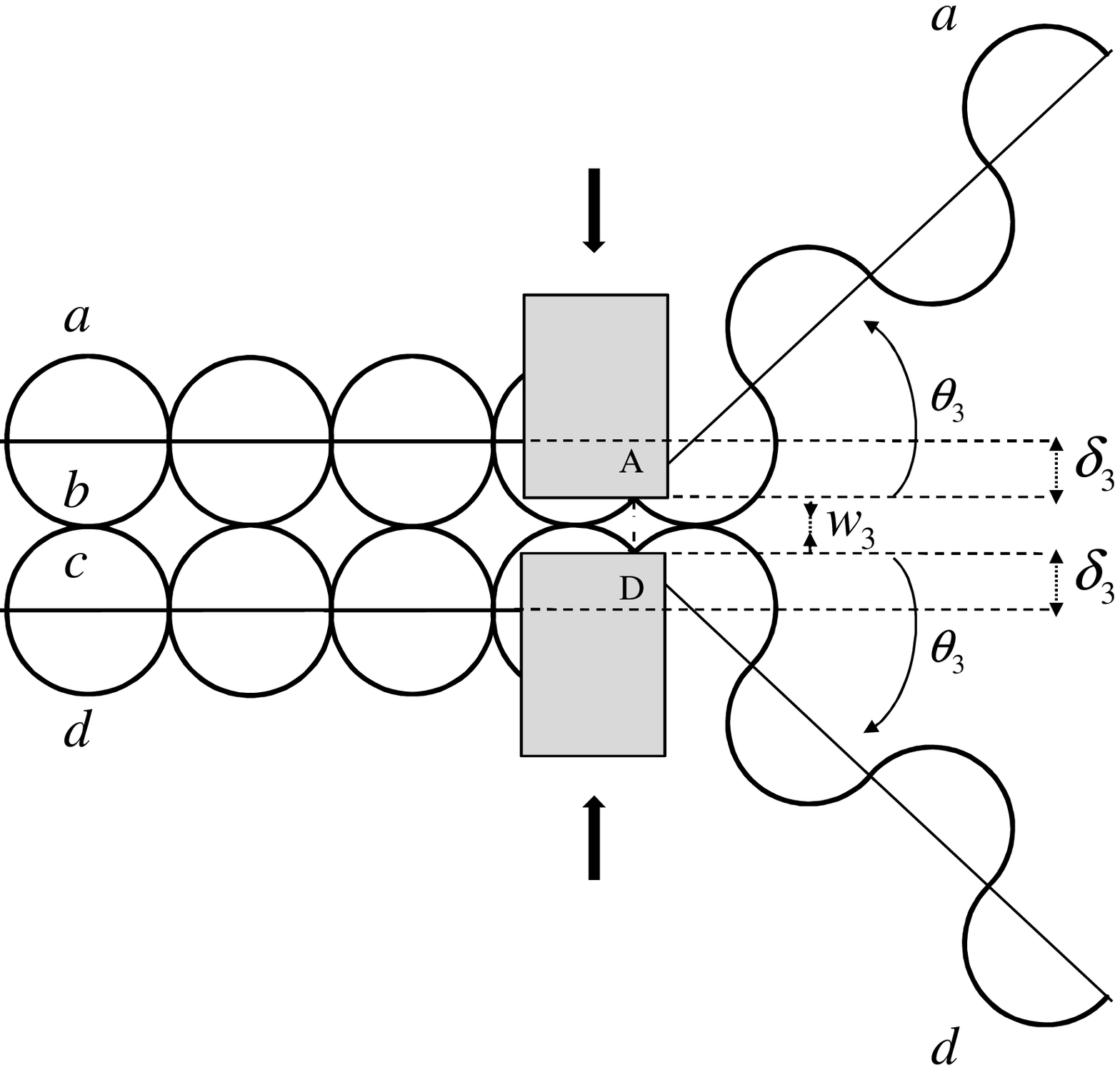}}
\caption{\footnotesize The deviation of the geodesics axis satisfies $\sin\theta_3={\delta_3\over r}$, when $2\delta_3+w_3=2r$.}\label{Fig04}
\end{minipage}
\end{figure}

As the slit becomes narrower, the geodesics $b$ and $c$  are totally reflected by the vertical side of the slit, meanwhile the geodesics $a$  and $d$ pass through the slit and get reflected by the horizontal side of the slit at the incident points $A$ and $D$ following the initial geodesics axis (as illustrated in Fig.\ref{Fig02}). The reflection of the geodesic $a$ (respectively $d$) at the incident point $A$ (respectively $D$) locally verifies the Fresnel law of reflection (see Fig.6, and explanation in ii)).
If the width of the slit becomes narrower, less that $2r$ (See Fig.\ref{Fig03} where $w_2<2r$), then the geodesics axis of the geodesics $a$ and $d$ are totally obstructed by the vertical side of the slit (see Fig.6, (c) and explanation in iii)). Therefore the geodesic $a$ (respectively $d$) is reflected by the horizontal side of the slit at the incident point $A$ (respectively $D$), and deviates its geodesic axis direction from horizontal (as illustrated in (c) Fig.6) under the assumption that if the initial geodesics axis are totally obstructed, the incident points $A$ and $D$ on the horizontal side of the slit become antipodal points for the new geodesics axis direction.

Using the Fresnel law of reflection at the incident point $A$ (respectively $D$), the deviation of the geodesics axis from horizontal satisfies the following equation as the slit gets narrower:
\begin{equation}\label{Dev}
\sin\theta={\delta\over r},\qquad\hbox{when}\quad 2\delta+w=2r.
\end{equation}
which yields
\begin{equation}\label{LDev}
\sin\theta=1-{w\over 2r}.
\end{equation}
where $\theta$ represents the angle of deviation of the geodesics axis of the reflected geodesic at the incident points $A$, respectively $D$, from horizontal (the horizontal side of the slit) and $\delta$ is the distance between the incident geodesics axis and the horizontal side of the slit such that $2\delta+w=2r$, where $w$ is the slit's width and $r$ the radius of the geodesics defined in (\ref{Rot}) and illustrated in Fig.2.

The narrower the slit is, the bigger the deviation of the geodesics axis is. Indeed, this can be understood straight forward from equality (\ref{LDev}), or by using the illustration in Fig.\ref{Fig03} and Fig.\ref{Fig04}, the equation (\ref{Dev}) for $\delta_3>\delta_2$ yields
\begin{equation}
{\sin\theta_3\over\sin\theta_2}={\delta_3\over \delta_2},
\end{equation}
thus for $\theta_i\in ]0,{\pi\over2}[$, $i=2,3$, we have $\theta_3>\theta_2$ for $\delta_3>\delta_2$. As the slit becomes narrower with a width $w_1=2r$, only two geodesics out of four initial geodesics pass thought the slit, and when the width $w_2$ (respectively $w_3$ ) of the slit becomes narrower, less that $2r$, the geodesics that pass through the slit are diffracted with a geodesics axis that makes an angle with the horizontal given by equation (\ref{LDev}).

\subsection{The Laser Beam Diffraction}

\begin{figure}[h!]
\begin{minipage}[t]{7cm}
\centering
\fbox{\includegraphics[width=5cm]{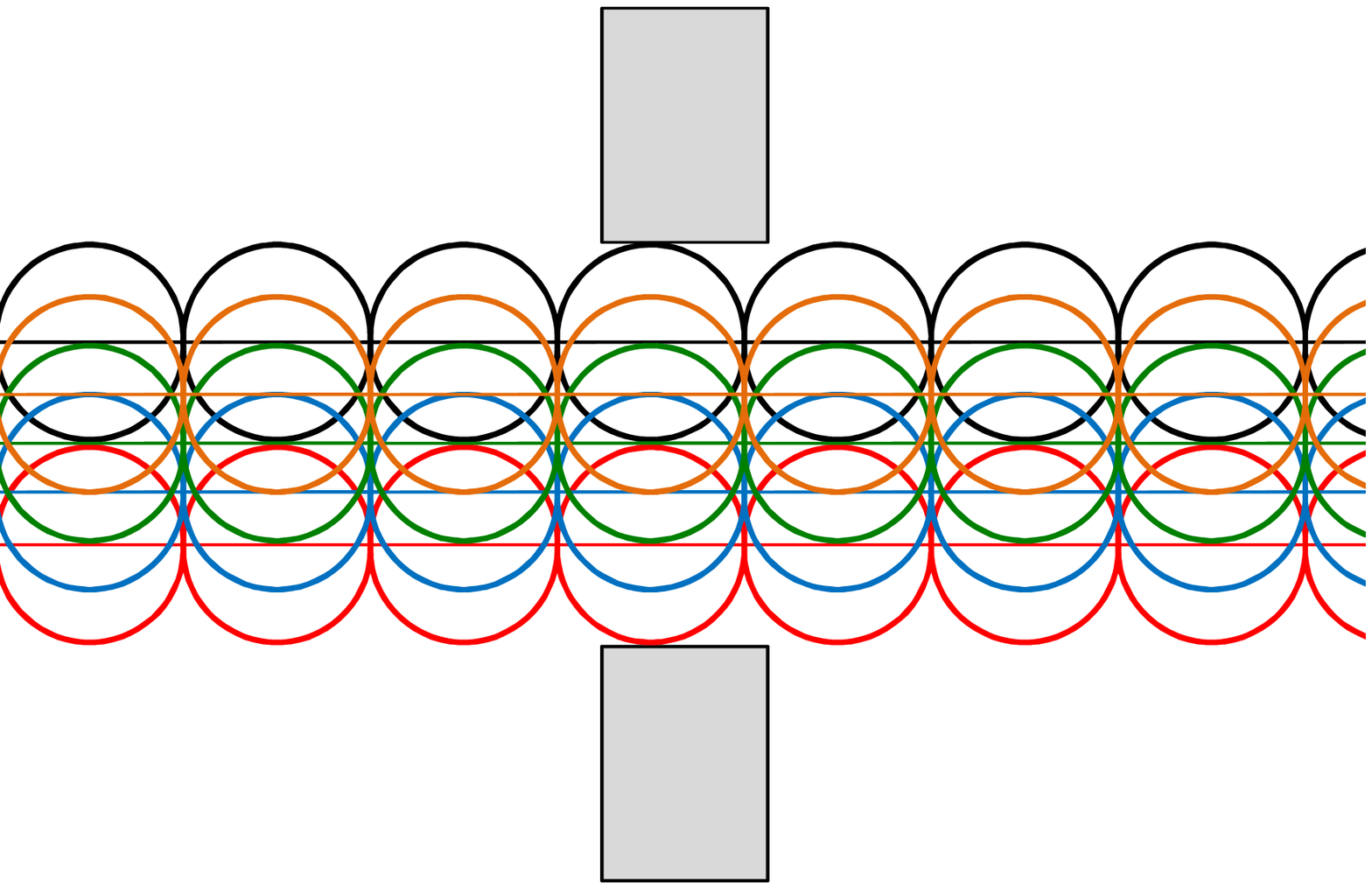}}
\caption{\footnotesize Illustration of  10 fluctuating geodesics that pass, from the left to the right, through a single slit of width $w_0=4r$. Each color represents a polarized geodesics given by (\ref{Rot}) in the same plane as the slit. Each color represents two fluctuating geodesics, one upper geodesic (the fluctuation starts up from the left), and one down geodesic (the fluctuation starts down from the left). The upper geodesic and the down geodesic are in anti-phase and they have the same geodesics axis. In this illustration $r=3.25mm$.}\label{Fig001}
\end{minipage}
\hspace*{\fill}
\begin{minipage}[t]{7cm}
\centering
\fbox{\includegraphics[width=5cm]{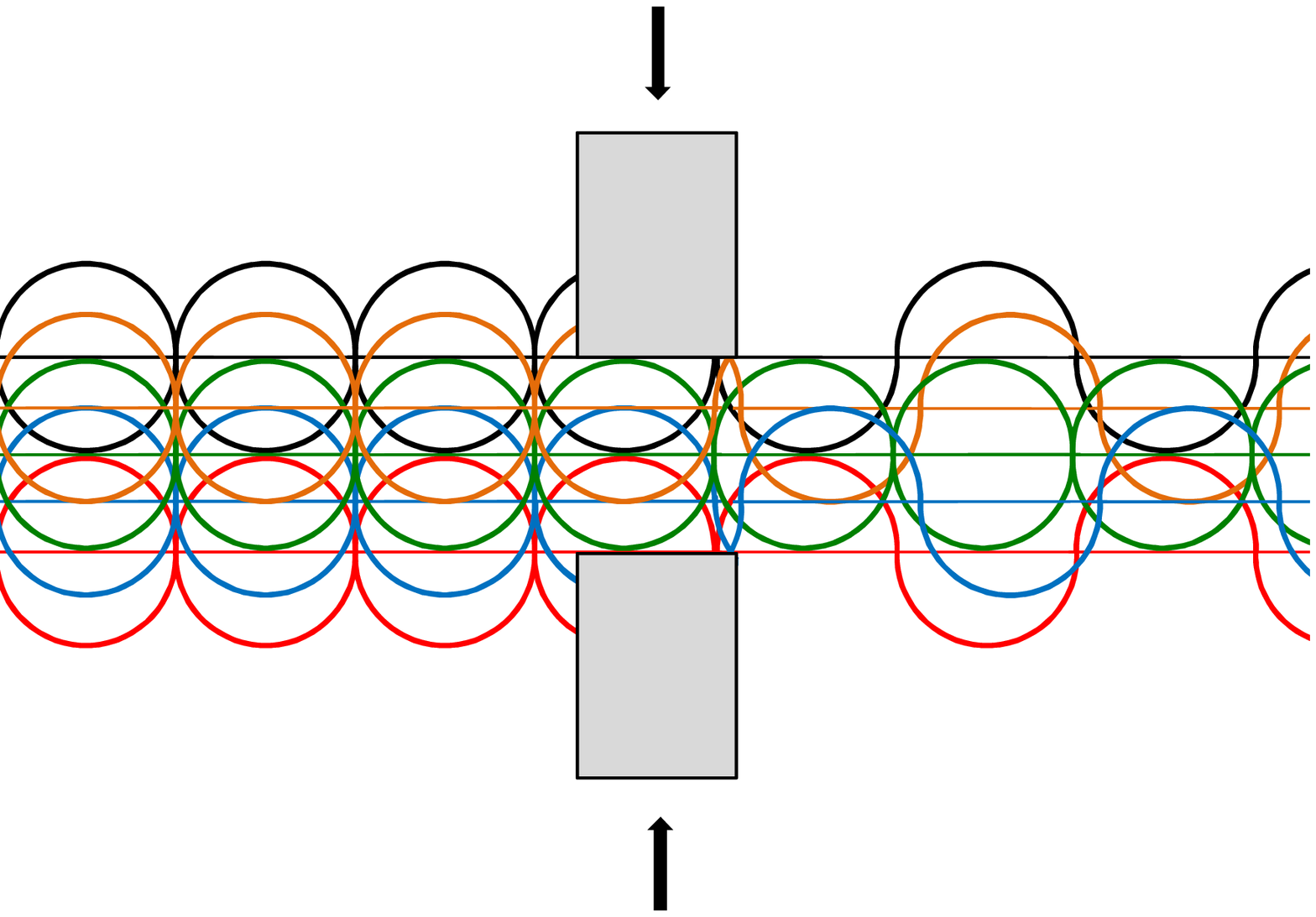}}
\caption{\footnotesize If the width of the slit $w_1$ is equal to $2r$. Then the lower black geodesic, the lower orange geodesic, the upper red geodesic and the upper blue geodesic  are obstructed by the vertical sides of the slit. The upper black geodesic and the lower red geodesic are reflected by the horizontal side of the slit and pass through the slit with the green geodesics to form a reduced spot in the detector screen.}\label{Fig002}
\end{minipage}
\end{figure}

\begin{figure}[h!]
\begin{minipage}[t]{7cm}
\centering
\fbox{\includegraphics[width=5cm]{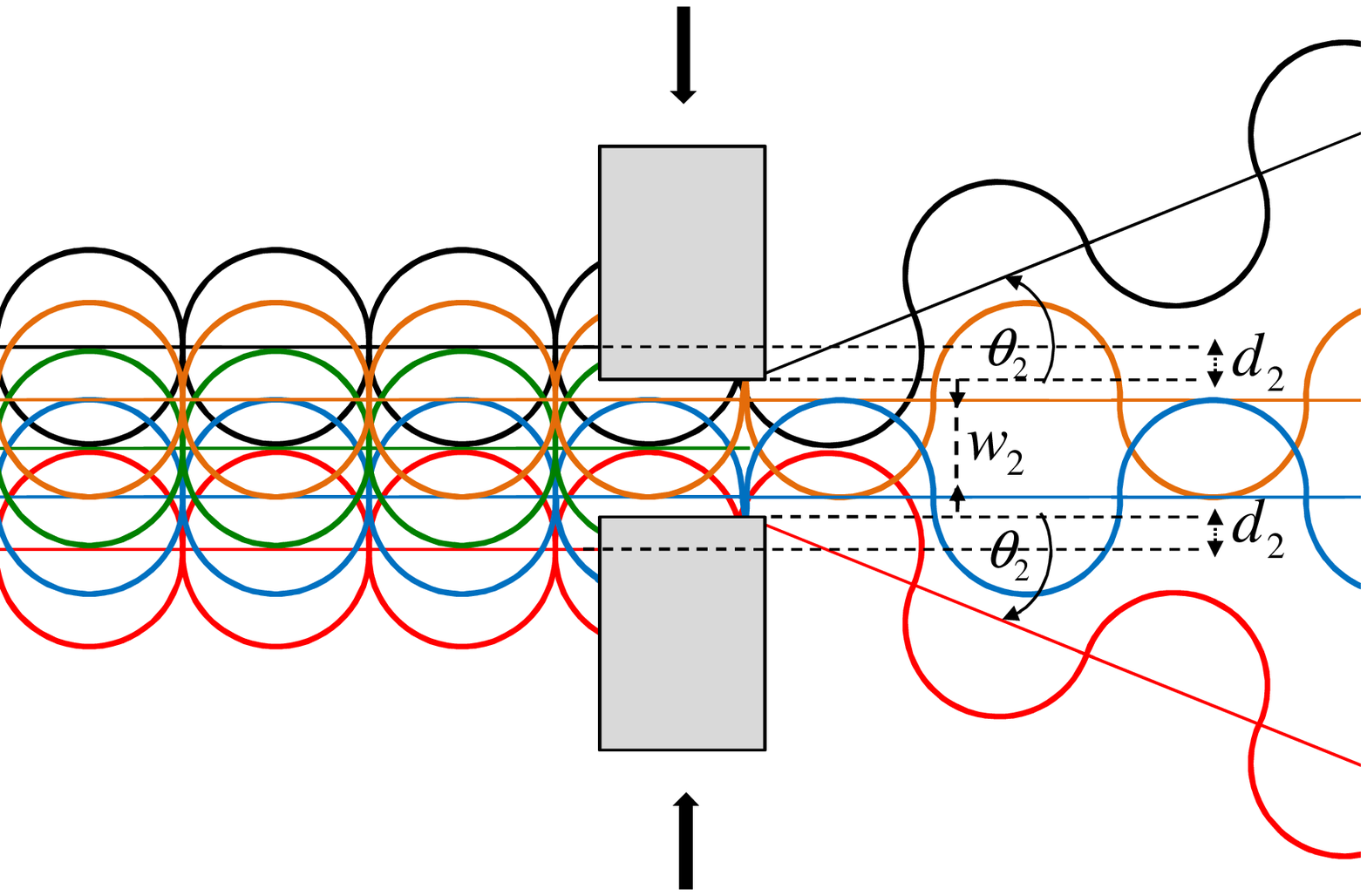}}
\caption{\footnotesize The width of the slit verifies $w_2<2r$, where $r$ is the radius of the geodesic (\ref{Rot}). Only 4 geodesics pass through, the upper black geodesic and lower red  geodesic are reflected by the horizontal slit's side. Their geodesics axis are deviated with an angle $\theta_2=\arcsin(\di{d_2\over r}$), since their overall incident directions are obstructed by the vertical slit's side.}\label{Fig003}
\end{minipage}
\hspace*{\fill}
\begin{minipage}[t]{7cm}
\centering
\fbox{\includegraphics[width=5cm]{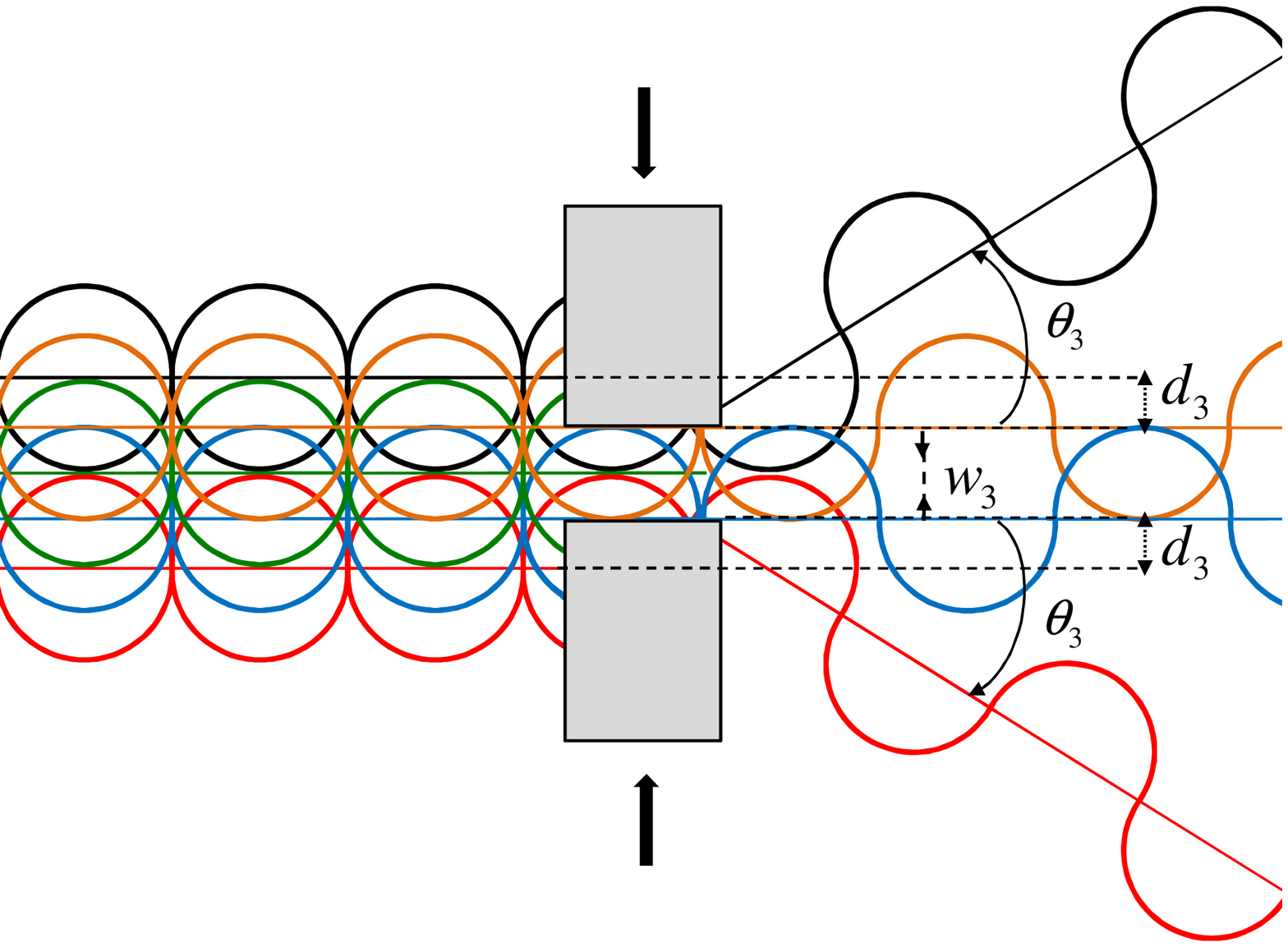}}
\caption{\footnotesize The width of the slit $w_3$ verifies $w_3<w_2<2r$, where $w_2$ is the width of the slit in Fig.\ref{Fig003}. The upper orange geodesic and the lower blue geodesic are reflected by the horizontal slit's side without deviation since their geodesics axis are not obstructed by the vertical slit. The upper black and lower red geodesics are reflected by the horizontal slit's side and the deviation of their geodesics axis increases together with the decreasing slit's width, $\theta_3=\arcsin(\di{d_3\over r}$) and $\theta_3>\theta_2$ for $d_3>d_2$.}\label{Fig004}
\end{minipage}
\end{figure}

For a small area concentrated with polarized geodesics defined by (\ref{Rot}) to form a laser beam, it is sufficient to illustrate the process with five pairs of geodesics concentrated in a small area (see Fig.\ref{Fig001}) relative to the wave length of light, where each color represents two anti-phase possible geodesics obtained by polarization of the geodesics (\ref{Rot}) in a given plane. The Fig.\ref{Fig001} illustrates a single slit through which passes a laser beam. As we reduce the slit's width, the spot in a distant detector screen is reduced since the number of fluctuating geodesics that pass through the aperture is reduced (only 6 fluctuating geodesics out of 10 incident fluctuating geodesics pass through the aperture in Fig.\ref{Fig002}). The spot of the laser beam in a distant detector screen is reduced together with the slit's width $w\in [2r, 4r[$. When the slit's width $w$ is strictly less than $2r$, the laser spot on the distant detector screen becomes wider and wider as the slit continues to be narrower and narrower (see Fig.\ref{Fig003}, and Fig.\ref{Fig004}). This is due to the deviation of the geodesics axis of the geodesics that bypass the slit's width whenever their incident geodesics axis is totally obstructed (as explained in Fig.\ref{Fig03} and Fig.\ref{Fig04}).
\pesp

\pesp

The simulation of a laser beam concentrated in a small area relative to the wave length of light in three dimensions by using the geodesics (\ref{Rot}), allows to understand the non intuitive behavior of light using a simple geometrical interpretation without involving the Heisenberg's uncertainty interpretation. More precisely, the spot of the laser beam in the distant detector screen gets narrower and narrower together with the vertical slit, and suddenly the spot in the detector screen changes and gets wider and wider horizontally as the vertical slit continues to get narrower and narrower.

If the vertical slit becomes narrower for a concentrated family of geodesics defined by (\ref{Rot}) in a small area relative to the wave length of light in three dimensions, the spot in a distant detector screen will get narrower together with the slit's width since the number of obstructed fluctuating geodesics by the slit's boundary increases until the width of the slit becomes relative to the half of the light wave length. Under this small width, the laser spot on the distant screen changes of behavior and  gets wider  horizontally as the slit continues to get narrower because of the total obstruction of the incident geodesics axis by the slit and the reflection of the fluctuating geodesics  by the slit's inner side (in Fig.\ref{Fig03} the inner side of the slit corresponds to the horizontal slit's side).

\section{Conclusion}

It is known that diffraction arises with all types of wave and its explanation can be found in Huygens' wave theory (\cite{HC},\cite{FA1}) (under the assumption of  Huygens-Fresnel principle). Observations of this phenomenon with many other types of physical systems (atoms, electrons, molecules, photons, protons, neutrons) around edges and barriers with slits, are considered as a pure manifestation of wave-like nature of the physical system. It is also known that geometrical optic failed to bring a rational explanation of the deviation of the light ray from rectilinear when it encounters an edge of a barrier or a narrow slit in a screen. This absence of consistent explanation of the light diffraction is maybe due to our misunderstanding of the space-time for the small scale world (\cite{BA}), or to our misunderstanding of the real nature of paths of least time for a physical system at the small scale world (quantum scale) that are shaped by the nature of the geodesics of the space-time.\pesp

Within this work, the diffraction of light is explained using an infinity of fluctuating geodesics (\ref{Rot}) that might be taken by photons in their quest of path of least time that minimizes the needed time to travel between two distant locations in an homogeneous space. The explanation of light diffraction completes the previous interpretation (\cite{BF}) of the interference pattern of Young's double-slit experiment. Indeed:\pesp

i) if the infinity of geodesics (\ref{Rot}) are considered as paths of least time for the physical system in an homogeneous space-time, the use of these fluctuating geodesics for the small scale world (the quantum scale) provides a consistent explanation of the diffraction of light, when it passes through a narrow slit, as a consequence of the reflection of these infinity of fluctuating geodesics on the edge of the narrow slit. This interpretation is consistent with the observed flaring of light when it encounters a narrow slit, and observation of these infinity of geodesics (\ref{Rot}) with small radius at the macroscopic scale reveals a straight line geodesic as path of least time.\pesp

ii) the superimposition of two copies of diffracted fluctuating geodesics (\ref{Rot}) when they pass thought narrow slits reproduces the interference pattern observed in  Young's double-slit experiment, and not only on the detector screen but also on the region between the slits and the detector screen (see \cite{BF} fo the highly reproducible simulations), which provides a rigorous explanation of the interference pattern without using the wave theory and formalism.\pesp

iii) the non intuitive behavior of the laser beam spot in the detector screen from narrower and narrower together with the slit's width to wider and wider as the  slit's width becomes smaller than a tiny slit's width find its interpretation using the infinity of geodesics (\ref{Rot}) with a consistent and a highly reproducible simulations. The geodesics (\ref{Geod}) that led to this rational interpretation were found in the simulation of an expanding Space-Time that expands via expansion of its basic elements. This simulation has shown in \cite{BP} that the geodesics in an homogenous and isotropic expanding space-time cannot be straight line geodesics, and the geodesics of such space-time are not only curved but also fluctuating, which attract our attention to relate the space-time print to explain the non intuitive behavior of elementary physical system in the quantum world.

The reproduction of interference pattern using only space-time geodesics reflects the print of the space-time characteristics rather than the real nature of the physical system and puts into question our interpretation of the nature of the physical system as a consequence of the observed diffraction phenomenon, or the observed interference phenomenon. The wave explanation of light diffraction is and remains a good approximation of the phenomenon, but it is not the only one and it is not sustainable as a determinant factor for the physical system nature since diffraction and interference can be explained with paths and trajectories. This new insight makes light particle-like nature conciliable with the so-called interference phenomenon.\pesp

 \end{document}